\documentclass[10pt]{elsarticlemod}

\usepackage{graphicx}
\usepackage{enumerate}
\usepackage{caption}
\usepackage{subcaption}
\usepackage{amssymb,amsmath}
\usepackage{bm}
\usepackage{rotating}
\usepackage{nicefrac}
\graphicspath{{figures/}}
\usepackage{color}
\usepackage{gensymb}

\begin{document}
%
\begin{frontmatter} 
\title{Numerical simulation of wave propagation and snow failure from explosive loading} 
\author[sfu]{Rolf Sidler}
\ead{rsidler@gmail.com}
\author[slf]{Stephan Simioni}
\author[imes]{J\"urg Dual}
\author[slf]{J\"urg Schweizer}
\address[sfu]{Department of Earth Sciences, Simon Fraser University, 8888 University Drive, BC V5A1S6 Burnaby,
Canada}
\address[slf]{WSL Institute for Snow and Avalanche Research SLF, Fl\"uelastrasse 11, 7260 Davos Dorf, Switzerland}
\address[imes]{Institute of Mechanical Systems, ETH Zurich, 8092 Z\"urich, Switzerland}

\begin{abstract}

Avalanche control by explosion is a widely applied method to minimize the avalanche risk to infrastructure in snow-covered mountain areas.
However, the mechanisms involved leading from an explosion to the release of an avalanche are not well understood.
Here we test the hypothesis that weak layers fail due to the stress caused by propagating acoustic waves. The underlying mechanism is that the stress induced by the acoustic waves exceeds the strength of the snow layers.
We compare field measurements to a numerical simulation of acoustic wave propagation in a porous material.
The simulation consists of an acoustic domain for the air above the snowpack and a poroelastic domain for the dry snowpack.
The two domains are connected by a wave field decomposition and open pore boundary conditions.
Empirical relations are used to derive a porous model of the snowpack from density profiles of the field experiment.
Biot's equations are solved in the poroelastic domain to obtain simulated accelerations in the snowpack and a time dependent stress field.
Locations of snow failure were identified by comparing the principal normal and shear stress fields to snow strength which is assumed to be a function of snow porosity.
One air pressure measurement above the snowpack was used to calibrate the pressure amplitude of the source in the simulation. 
Additional field measurements of air pressure and acceleration measurements inside the snowpack were compared to individual field variables of the simulation.
The acceleration of the air flowing inside the pore space of the snowpack was identified to have the highest correlation to the acceleration measurements in the snowpack.

\end{abstract}


\end{frontmatter}

\section{Introduction}

During the last decades the number of people living and recreating in, or travelling through mountainous terrain has substantially increased.
To ensure the reliability of infrastructure extensive engineering works such as supporting structures and snow sheds have been built to prevent damages due to large avalanches. 
Whereas these permanent protection measures are highly effective, they are also costly. Therefore, less expensive temporary preventive measures have become increasingly popular over the last decade. In particular, artificial avalanche release by explosion is among the key preventive measures. The aim is to trigger avalanches when their size is still small enough to not cause any damage and no people are exposed in the path of the avalanche \citep{mcclung:2006}. 

Releasing avalanches with explosives by hand or helicopter charging is, however, limited to locations or weather conditions allowing tolerably save access for avalanche control personnel. This limitation has been overcome by fixed avalanche control installations which trigger avalanches by the effect of explosions and allow remote operation even under most adverse weather conditions or during nighttime.
The basic physical mechanisms that cause slab avalanches to release from explosives, and other causes, are well known and have been used to choose optimal locations of blast installations for years. What is lacking is a quantitative model incorporating the ``known'' physics associated with initiating failure of slab avalanches that can be used to examine the processes, improve understanding of the physical processes and make predictions that can be tested in the future.

Historically, research on avalanche control has been focused on experimental evidence of waveforms, charge type and placement to support the work of avalanche control operations \citep{gubler:1976,mellor:1973,ueland:1993,bones:2012,binger:2015}. The most extensive measuring campaigns were performed by \citet{gubler:1977}. However, many of the more recent studies focused on small range effects \citep{bones:2012,wooldridge:2012,johnson:1993}. A more detailed review of the past research within snow and explosions is given by \citet{simioni:2015}. 
A model considering the porous character of snow based on Biot's \citeyearpar{biot:1962} equations has been proposed by \citet{johnson:1982}, but has rarely been applied to snow since \citep{albert:2013}. A mixed stress-energy failure criterion including simplified effects of explosive loading was developed by \citet{cardu:2008}. It is only recently that numerical tools are used to support a theoretical framework on the physical mechanisms that lead to the release of an avalanche.  \citet{miller:2011} considered the non-linear effects of an explosion and non-linear compaction of the snowpack for close ranges using the finite element method. 

Here we compare the measurements from field experiments on the wave propagation caused by an explosion to the results of a numerical simulation considering the porous character of snow. 
We tested the hypothesis that the stresses induced by the acoustic wave propagating through the snowpack locally exceed the snow strength and lead to failure.
In the winter 2013-2014 we performed multiple field experiments with avalanche control explosives triggered at different elevations above the snow surface and measured the air pressure above the snowpack as well as acceleration at different depths within the snowpack and distances from the point of explosion \citep{simioni:2015}. In addition, we recorded weak layer failure with cameras.

In the following we describe the numerical model that was used to perform the simulations. We focus on a specific experiment as a showcase for the test series, build a layered porous model for the prevailing snowpack, and evaluate the numerical results toward measured air pressure and acceleration. Finally, we compare locations where the stress in the numerical simulation exceeds the strength of the snowpack to the observed locations in the field.

\section{Methods}

\subsection{Field experiment}
\begin{figure}
	\centering
	\includegraphics[width=1\textwidth]{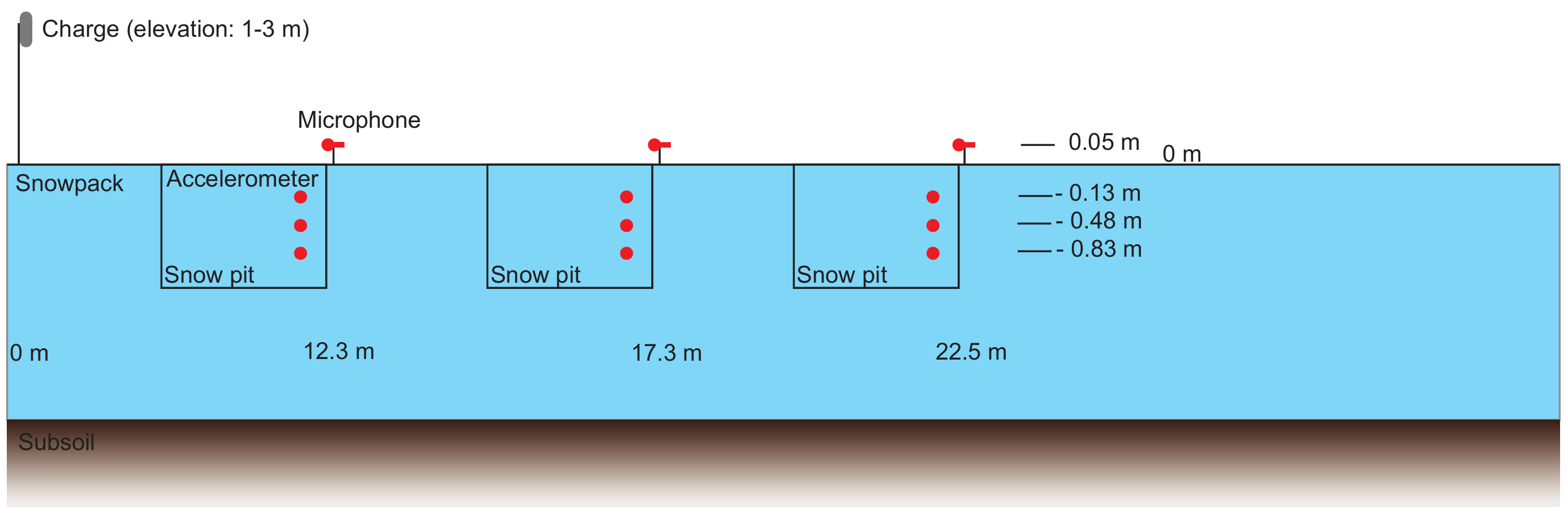}
	\caption{\label{fig:layout} Longitudinal section of the measuring layout of the experiments from 27 February 2014 \citep{simioni:2015}.}
\end{figure}

\begin{figure}
\centering
\includegraphics[width=1\textwidth]{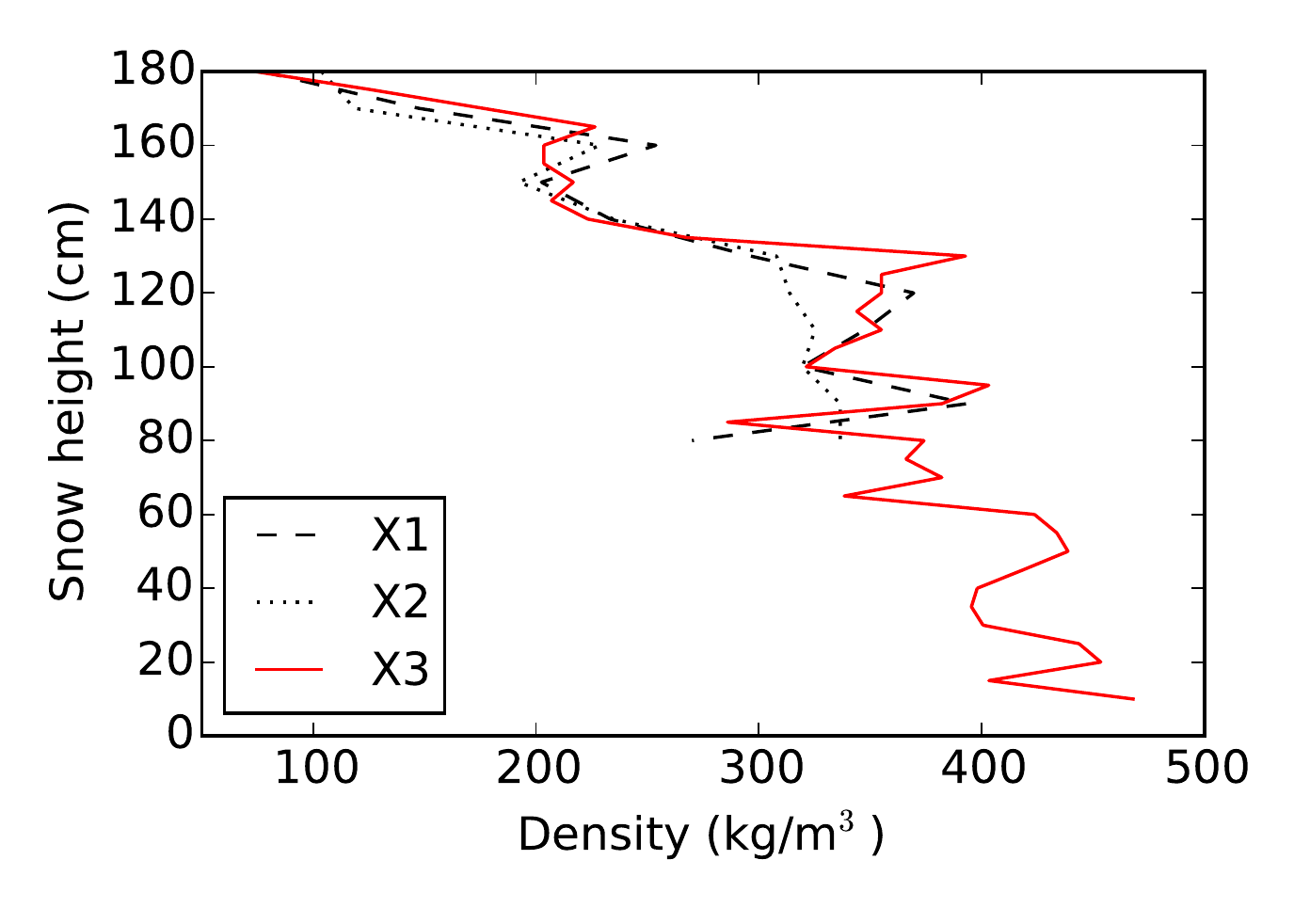}
\caption{\label{fig:snow-profile} Density profiles measured with the capacitive Denoth probe in the three pits at distances of 12, 17 and 22 m from the point of triggering.}
\end{figure}

We chose the first experiment from a day with eight experiments on 27. February, 2015 as a showcase to compare with the numerical results.
The geometry of the experiment is shown in Figure \ref{fig:layout}.
A 4.3~kg explosive charge was taped to a wood stick and placed 1~m above the snow surface.
Three snow pits were excavated 12.3~m, 17.3~m, and 22.5~m horizontal distance from the point of the explosive charge. Microphones were placed 0.05~m above the undisturbed snow surface next to the snow pits.
Three accelerometers were installed in cavities at 0.13~m, 0.48~m, and 0.83~m below the snow surface in each snow pit\citep{simioni:2015}. Special care was given to fit the diameters of the horizontal holes exactly with the diameters of the accelerometers to warrant the coupling of the sensors to the snowpack.
Snowpack failure was recorded with compact cameras \citep{simioni:2015}
%
%
The snowpack on the investigated day was 187~cm deep and consisted of a 45 cm thick layer of recently deposited snow (consisting of decomposing and fragmented precipitation particles) including two melt-freeze crusts above a well-consolidated base. The base was composed of layers of small rounded grains interspersed with several melt-freeze crusts and ice layers above hard layers of faceted crystals near the bottom of the snowpack. A potential weak layer was identified at a height of 85 cm from the ground. 
The snowpack was still dry but relatively warm with a minimum temperature of -1\degree C. The point snow stability based on the snow profile was rated as good \citep{schweizer:2001}. An extended column test  \citep{Simenhois:2009} indicated that the potential weak layer was very hard to trigger as it was buried below a 1~m thick well consolidated slab.
The densities obtained by capacitive measurements in the three snow pits are shown in Figure \ref{fig:snow-profile} \citep{denoth:1989,eller:1996}.
To localize weak layer failure during the experiments, compact cameras were installed in each snow pit and recorded the pit wall during the explosion. The single video stills allowed to visually identify weak layer failure due to movement of the snowpack overlaying the weak layer \citep{simioni:2015}.
%

\subsection{Numerical model}
Seasonal snow is a highly porous material with air often taking up the larger part of the volume. 
\citet{johnson:1982} showed that Biot's \citeyearpar{biot:1956} theory for wave propagation in porous materials can be successfully applied to snow.
Acoustic wave propagation in such porous materials is characterized by the presence of a compressional and a shear wave in the ice skeleton and an additional second compressional wave that is propagating in the pore fluid that is also called the ``slow'' wave. Due to the high porosity and the proportions between material properties this second compressional wave mode is propagating in snow \citep{oura:1952,ishida:1965} and can be recorded. This stands in contrast to other natural porous materials as, for example, sediments where the wave is diffusive and cannot be recorded.
The energy dissipation mechanism in Biot's theory is physically modeled by the viscosity of the pore fluid which is moved against the skeleton as acoustic waves propagate through the material.
Biot's model also accounts for the interaction between waves propagating in the porous frame and in the pore space of the material. 

To simulate acoustic wave propagation in snow we use a pseudo-spectral approach which is known to be accurate and efficient \citep{boyd:2001}.
We use the algorithm of \citet{sidler:2010a} where the simulation consists of an upper acoustic domain that is connected to a poroelastic domain by a wave field decomposition to account for the boundary conditions at the interface \citep{gottlieb:1982,carcione:1991b,tessmer:1992}.
For this study, the acoustic domain represents the air above the snowpack and the lower domain, where Biot's \citeyearpar{biot:1962} differential equations are solved, represents the snowpack.

Interfaces of porous materials are not uniquely defined and have one degree of freedom that can be interpreted as the connection between the pore space \citep{deresiewicz:1963}. 
The connectivity of the pore space is expressed with the so called surface flow impedance that is zero for open pores and infinite for closed pore interfaces.
For natural occurring materials the pore space is mostly connected and open pore boundary conditions apply. However, a coating on the surface, a deposit in the pore space or a mismatch of the pore throats between two porous materials or layers with different characteristics can lead to decreased connectivity.
For the snow surface we assume open-pore type boundary conditions, where the air in the pore space is fully connected to the air above the snowpack and the pore spaces of adjacent layers are fully connected. 

The snowpack is considered two-dimensional in the model. The total size of the acoustic domain is 29~m in horizontal direction by 20~m in vertical direction.
The poroelastic domain has the same length in horizontal direction, but is only 3.5~m in vertical direction.
The acoustic domain of the model consists of 185 grid nodes in vertical direction and 725 grid nodes in horizontal direction. The poroelastic domain has the same number of grid nodes in the horizontal direction and 147 grid nodes in vertical direction.
Due to the use of a Fourier operator the grid nodes are equally spaced at 0.04~m in horizontal direction. 
In vertical direction the spacing is irregular due to the Gauss-Lobatto points of the Chebyshev operator
and a consequent grid stretching \citep{tessmer:1994,peyret:2002}. 
Close to the interface the grid nodes are more densely spaced but almost regularly spaced at 0.04 m throughout most of the porous domain. 

The source was placed 4~m from the left boundary of the domain 1~m above the snow-air interface.
A total of $5.6 \times 10^5$ time steps with a length of $2 \times 10^{-7}$~s were computed, which corresponds to a total length of 0.112~s for the entire simulation.

A limitation in the presented simulation is that the non linear effects present in the vicinity of the explosion are not considered. These non-linear effects are believed to no longer be of relevance at the distances considered in this study. As the simulation also models the early part of the experiment some adjustments of the results of the simulation are necessary. The most significant adjustments and limitations of this simulation are:
\begin{enumerate}[i]
\item To account for the supersonic wave velocity of the shock wave in the vicinity of the explosion we have adjusted the timing of the simulation to the measurements of the arrival of the direct air wave at the pressure receiver in the air closest to the explosion.
\item To account for the unknown pressure amplitude at the point of the explosion we have scaled the the entire simulation to the pressure of the direct air wave measured at the closest pressure sensor in the air. This is appropriate as the applied simulation is based on linear equations. The relative amplitude of field variables in the simulation does not depend on the amplitude of the source. Therefore it is possible to compute a simulation with a random source amplitude and scale the entire simulation with the actual pressure of the source or with a measurement of any of the field variables at any location in the simulation.
\item To account for the unknown source waveform and changes in the waveform due to non-linear effects we have chosen the Friedlander wavelet as the source waveform. This is the simplest form to express a blast wave \citep{friedlander:1946}.
\item To account for the effects of coupling between the acceleration of the pore fluid (air inside the snowpack) and the accelerometers we have used a simple scaling of the simulated fluid acceleration by a scaling factor.
\item The simulation does not make any adjustments at locations where snow failure is predicted. 
\end{enumerate}

\subsection{Snowpack model}

Ten properties of the porous material are required to solve Biot's equations.
The ice skeleton of the snow is defined by the bulk modulus of the frame material $K_s$, the bulk modulus of the matrix $K_m$, and the shear modulus $\mu_s$.
The pore fluid is characterized by the fluid bulk modulus $K_f$ and the viscosity $\eta$ of the pore fluid.
Additionally, the densities of the solid and fluid materials $\rho_s$ and $\rho_f$ have to be known.
The geometry of the pore space is defined by the porosity $\phi$, the permeability $\kappa$, and the tortuosity $\mathcal{T}$.
It is often not possible to measure all these properties at the required spatial resolution. 
However, the properties of snow are interrelated and {\it a priori} information, geometrical considerations, and empirical relationships can be used to estimate unknown snow properties from snow porosity or density \citep{sidler:2015}. 
The relations to obtain the porous properties of snow from its porosity are summarized in Table~\ref{tab:snowmodel}.

\begin{table}
\caption{Material properties for a Biot-type porous snow model as a function of porosity $\phi$ \citep{sidler:2015}.}
\begin{center}
\begin{tabular}{lll}
\hline
Porous frame & \\
\hline
 	frame bulk modulus	& $K_{\rm s}$ (GPa) 	& 10  \\
 	matrix bulk modulus	& $K_{\rm m}$ (GPa)	& $K_{\rm S} (1-\phi)^\frac{30.85}{(7.76-\phi)}$ \\
 	shear modulus		& $\mu_{\rm s}$ (GPa)	& $  \frac{3}{2} \frac{K_{\rm m} (1-2 \nu)}{1+\nu}$ \\
	density			& $\rho_s$ (kg/m$^3$)		& $(1-\phi) \cdot 916.7 $ \\
	permeability		& $\kappa$ (m$^2$)		& $0.2 \frac{\phi^3}{(\text{SSA})^2 (1-\phi)^2}$ \\ 
	tortuosity			& ${\cal T}$			& $\frac{1}{2} \left( 1+\frac{1}{\phi} \right)$ \\
	Poisson's ratio		& $\nu$				& $\nu = 0.38 - 0.36 \, \phi$ \\
	specific surface area & $\text{SSA}$	(m$^{2}$/kg)		& ${ \rm SSA} =  -30.82 \frac{{\rm  m^2}}{{\rm  kg}} \ln(1-\phi) -17.93 \frac{{\rm  m^2}}{{\rm  kg}}$ \\							
\hline
Pore fluid & \\
\hline
	density		& $\rho_{\rm f}$ (kg/m$^{3}$)	& 1.29 \\
 	viscosity		& $\eta_{\rm}$ (Pa s)		& $1.7 \times 10^{-5}$  \\
	bulk modulus	& $K_{\rm f}$ (Pa)			& $1.4 \times 10^{5}$  \\
\hline
\end{tabular}
\end{center}
\label{tab:snowmodel}
\end{table}%

\begin{figure}
\centering
\includegraphics[width=1\textwidth]{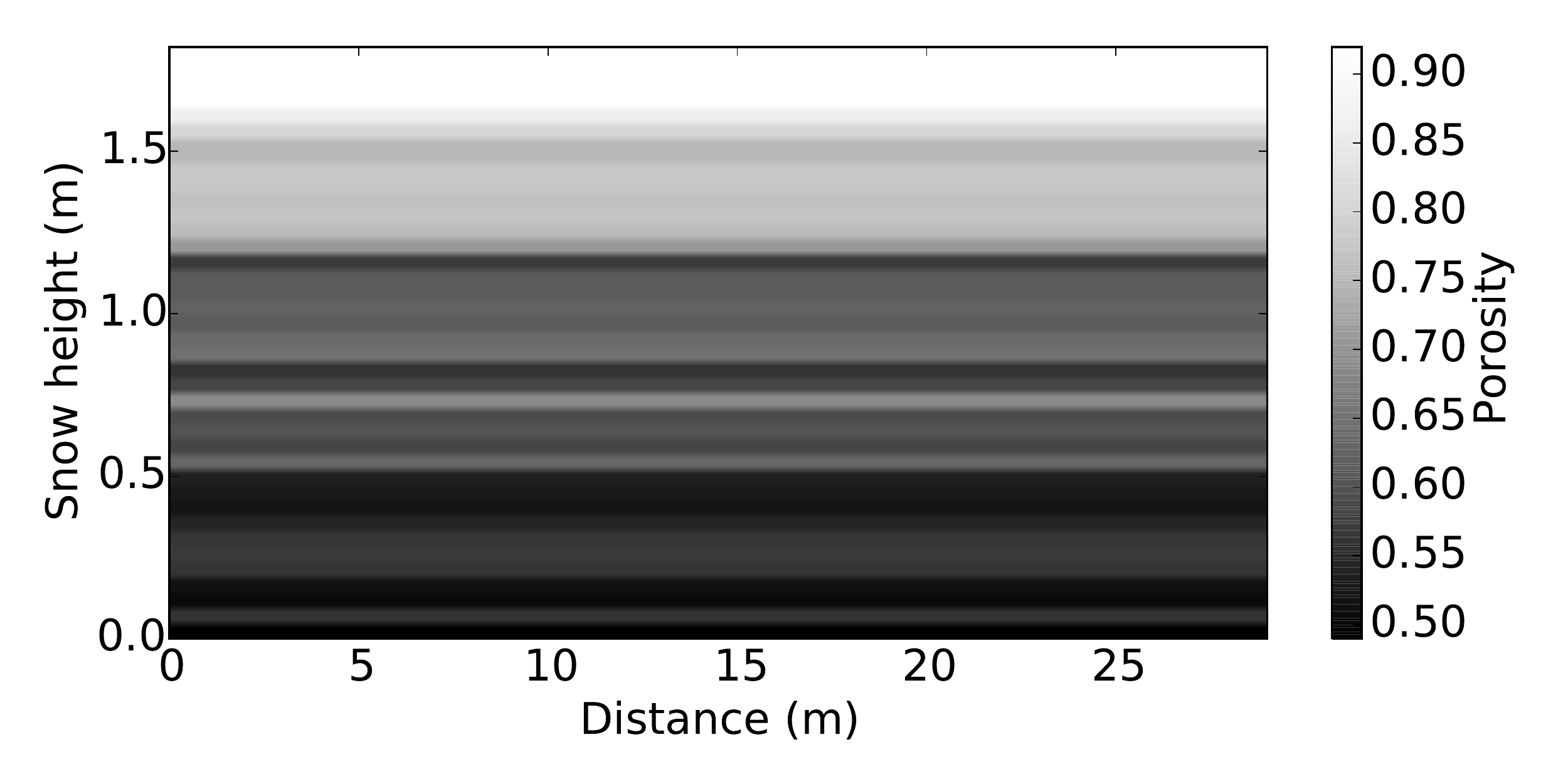}
\caption{\label{fig:porosity} Porosity distribution for the numerical model based on the snow profile in Hole 3.}
\end{figure}

The density profiles in the three snow pits show a spatially uniform distribution of the snowpack and we use a horizontally layered snowpack model for the simulation. 
Our density model is based on the density profile from the snow pit at 22~m horizontal distance from the explosion labeled X3 in Figure \ref{fig:snow-profile} and is shown in Figure \ref{fig:porosity}.
A porosity model is obtained from this density model by assuming that the pore space is completely filled with air in dry snow. Based on the porosity model the remaining properties for the porous model are derived according to the relations shown in Table~\ref{tab:snowmodel}.

\subsection{Source characterization}

The strong overpressure originating from an explosion leads to non-linear effects that are not covered by our numerical simulation. 
For such high pressures, the bulk modulus of the air is a function of the pressure amplitude and the temperature difference before and after the shock front \citep{cooper:1996}.
The higher bulk modulus of air under higher pressure and temperature leads to shock waves propagating faster than sound waves. 
Moreover, as parts of the waveform with higher amplitude propagate faster than those with lower amplitude the wave front steepens up during propagation.  
When the shockwave is reflected at the interface between the air and the snowpack the incident and the reflected wave have positive interference and consequently a higher velocity in the vicinity of the interface that can lead to the formation of a so called ``mach stem'' \citep{mach:1878}.

\begin{figure}
\centering
\large a) 
\begin{minipage}[t]{0.45\textwidth}
\vspace{-10pt}
\includegraphics[width=1\textwidth]{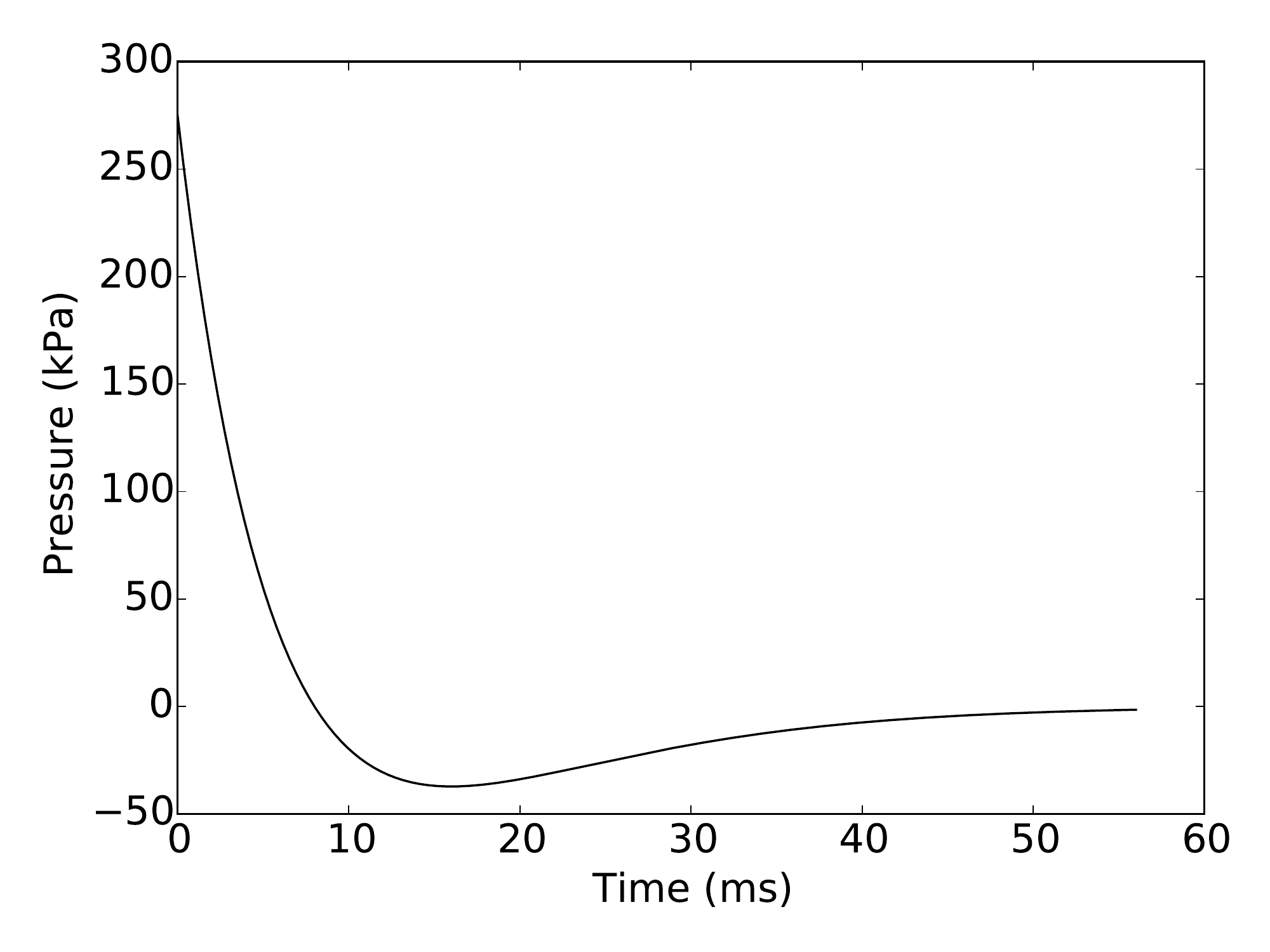}
\end{minipage}
\begin{minipage}[t]{2mm}
\large b) 
\end{minipage}
\begin{minipage}[t]{0.45\textwidth}
\vspace{-10pt}
\includegraphics[width=1\textwidth]{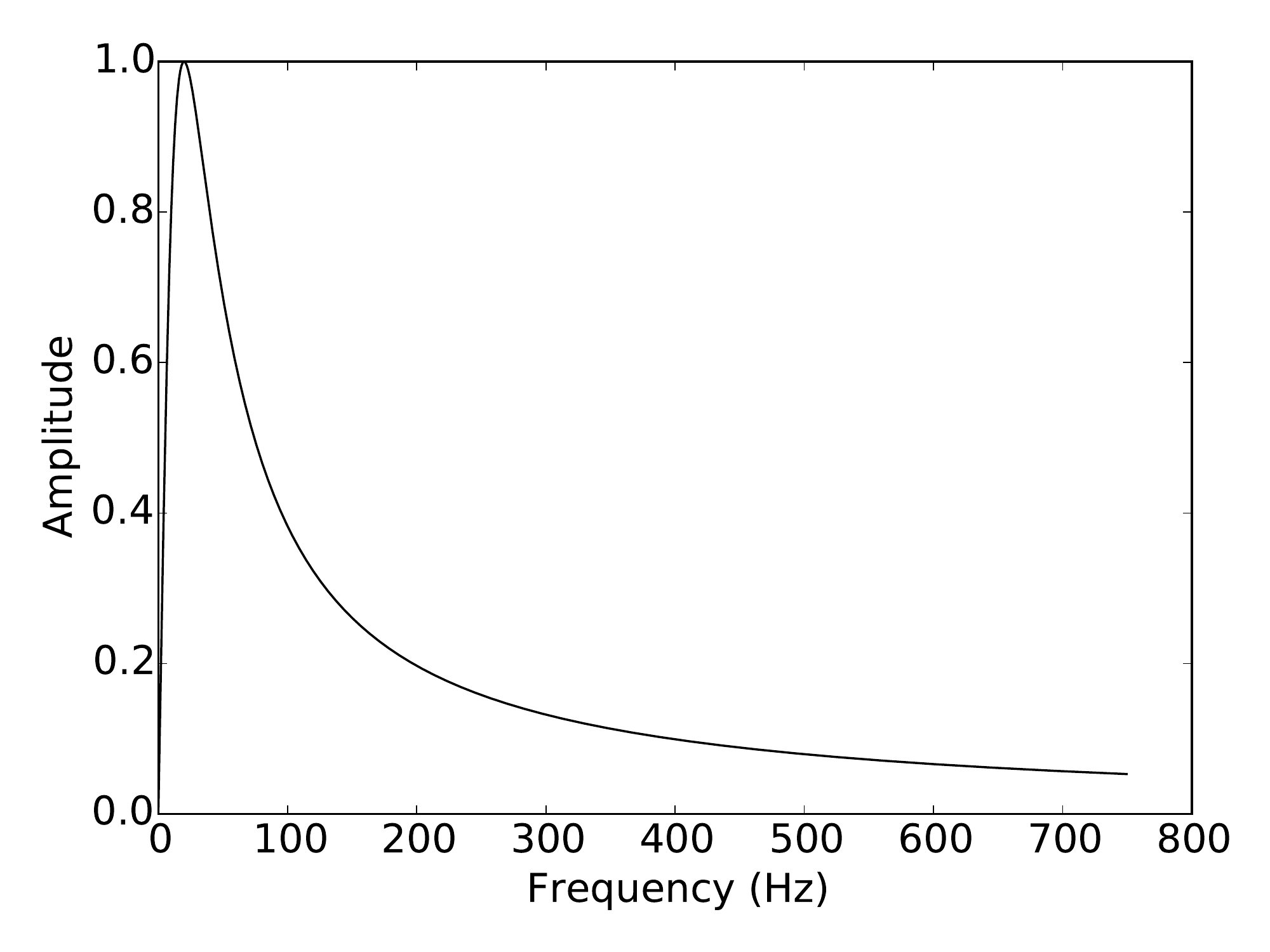}
\end{minipage} \\ 
\caption{\label{fig:source-wvlt} a) Friedlander wave form and b) frequency spectrum.}
\end{figure}

However, these effects are present only in the vicinity of the snowpack. The air pressure is supposed to decay due to non adiabatic effects and geometrical spreading. Therefore, at larger distances from the point of explosion, linear equations can be used to predict wave propagation.
Using an elastic approach, it is not possible to characterize the amplitude and waveform from the energy content and the type of chemical reaction of the explosive.
Instead, the waveform is estimated from pressure recordings to be of the form of a Friedlander wavelet, which is widely used for this purpose. 
This waveform is then scaled to fit the recorded pressure at the microphone that was placed at 12.3 m distance from the point of explosion.
A similar approach was used by \citet{albert:2013} who also used the recording of a receiver to characterize the waveform of the elastic equivalent source.
The Friedlander wavelet and its frequency content used in the numerical simulation are shown in Figure \ref{fig:source-wvlt}.

\subsection{Simulated snow failure locations}

\begin{figure}
\centering
\large a) 
\begin{minipage}[t]{0.45\textwidth}
\vspace{-10pt}
\includegraphics[width=1\textwidth]{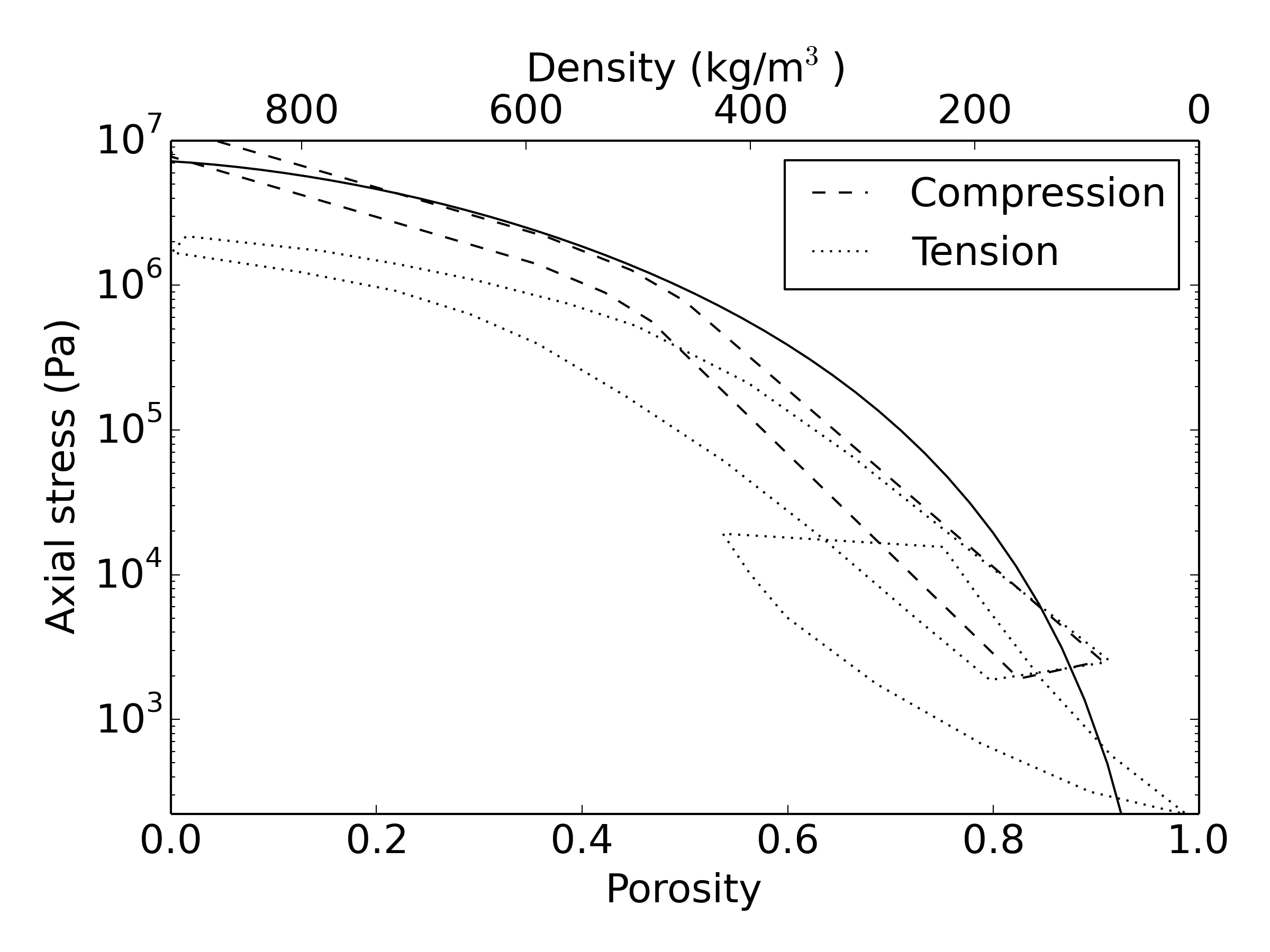}
\end{minipage}
\begin{minipage}[t]{2mm}
\large b) 
\end{minipage}
\begin{minipage}[t]{0.45\textwidth}
\vspace{-10pt}
\includegraphics[width=1\textwidth]{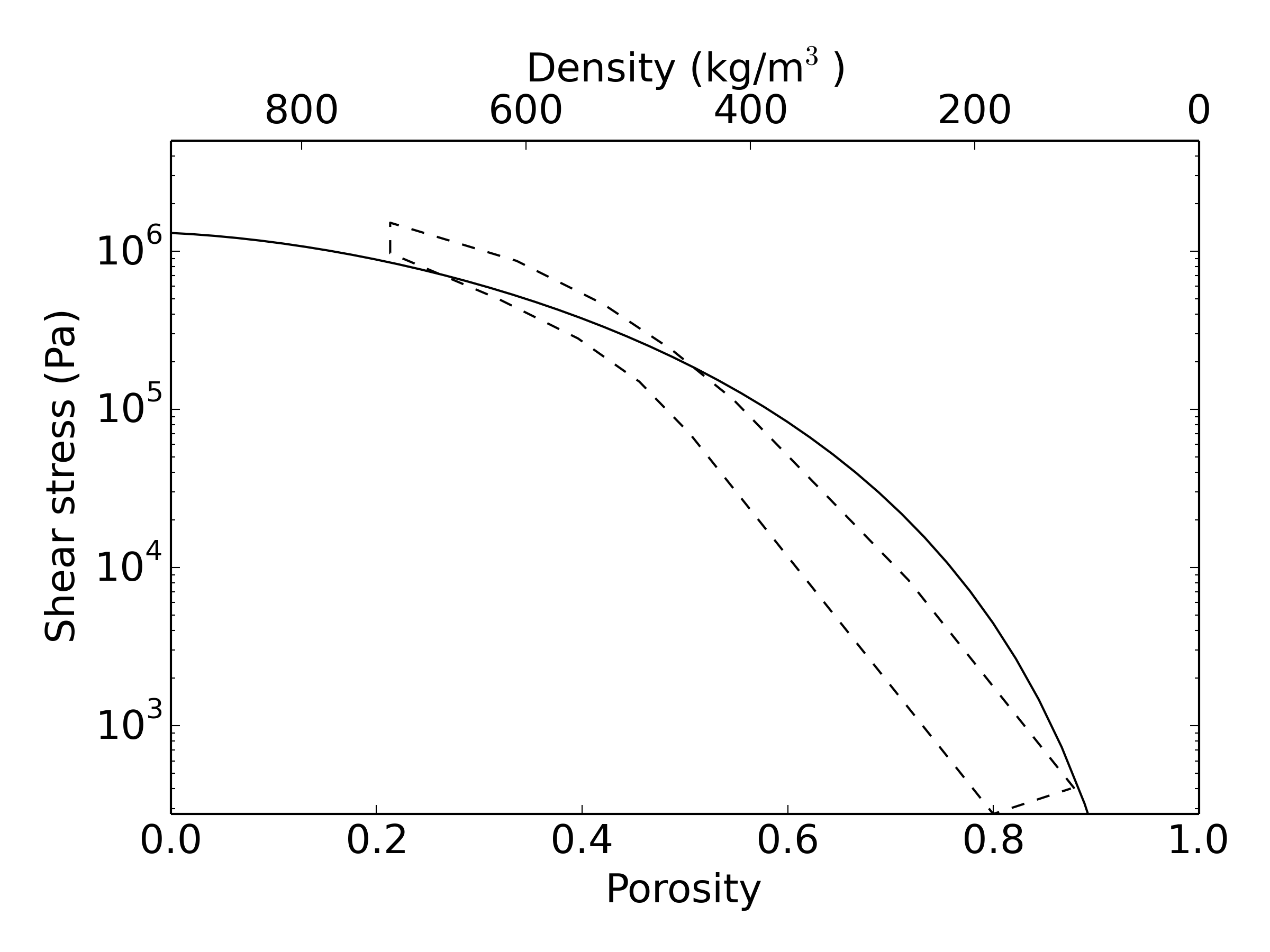}
\end{minipage} \\ 
\caption{\label{fig:strength} Maximum a) compressional and b) shear strength of snow. The black lines denote the strength to uniaxial and shear stress, respectively. For comparison, maximum strength of compilations from \citet{mellor:1975} and \citet{shapiro:1997} are shown.}
\end{figure}

To evaluate the locations where the propagating wave field leads to a failure of the snowpack we compare the stress field of the numerical simulation to the compressional and shear strength of snow.
In general, the nature of failure depends on the loading conditions. For brittle failure, \citet{mellor:1975} suggested to use a fraction of the snow's Young's modulus to define the maximum stress snow may withstand before it starts to fail. Here we use a fraction of $1 \times 10^{-3}$ of the bulk modulus and a fraction of $0.5 \times 10^{-3}$ of the shear modulus to define the strength.
The corresponding shear and compressional strength as a function of porosity based on these fractions and the relationships from Table~\ref{tab:snowmodel} are shown in Figure~\ref{fig:strength} and are compared to the compilation of snow strength measurements presented by \citet{mellor:1975} and \citet{shapiro:1997}.

The absolute values of the field variables were scaled to fit the pressure recording of the microphone at a horizontal distance of 12.3~m from the point of the explosion to evaluate the snow failure locations in the numerical simulation.
As the underlying equations are linear the scaling of the simulation can be performed by a simple multiplication with a scaling factor.
The factor itself depends on the amplitude of the source in the simulation and can be randomly chosen.

The maximum principal normal and shear stresses for all grid nodes in the simulation are computed from snapshots of the stress tensor every 0.4~ms for the whole length of the simulation.
The maximum principal normal stress $\sigma_{pn}$ and the principal shear stress $\tau_{p}$ can be obtained from the stress tensor $\sigma$ as 

\begin{equation}
\sigma_{pn} = \frac{\sigma_{xx} - \sigma_{yy}}{2} +\tau_{p}, 
\end{equation}
and
\begin{equation}
\tau_{p} = \sqrt{ \sigma_{xy}^2 +\left( \frac{\sigma_{xx} - \sigma_{yy}}{2} \right)^2 } ,
\end{equation}

where the first indices $x$ and $y$ indicate a plane normal to the corresponding coordinate axis and the second indices denotes the direction in which the stress acts \citep{jaeger:2007}.

The individual snapshots are then evaluated for locations where principal normal and shear stresses exceed the maximum strength of the snow model. These maximum strengths are computed from the porosity model.
Based on the porosity model matrix the bulk and the shear moduli for all grid nodes are computed using the equations shown in Table \ref{tab:snowmodel}.
Locations where the simulated stresses have exceeded the computed strengths of the snowpack in one or more snapshots are considered locations of snow failure.

\section{Results}

\subsection{Air pressure}

\begin{figure}
\centering
\large a) 
\begin{minipage}[t]{85mm}
\vspace{-10pt}
\includegraphics[width=1\textwidth]{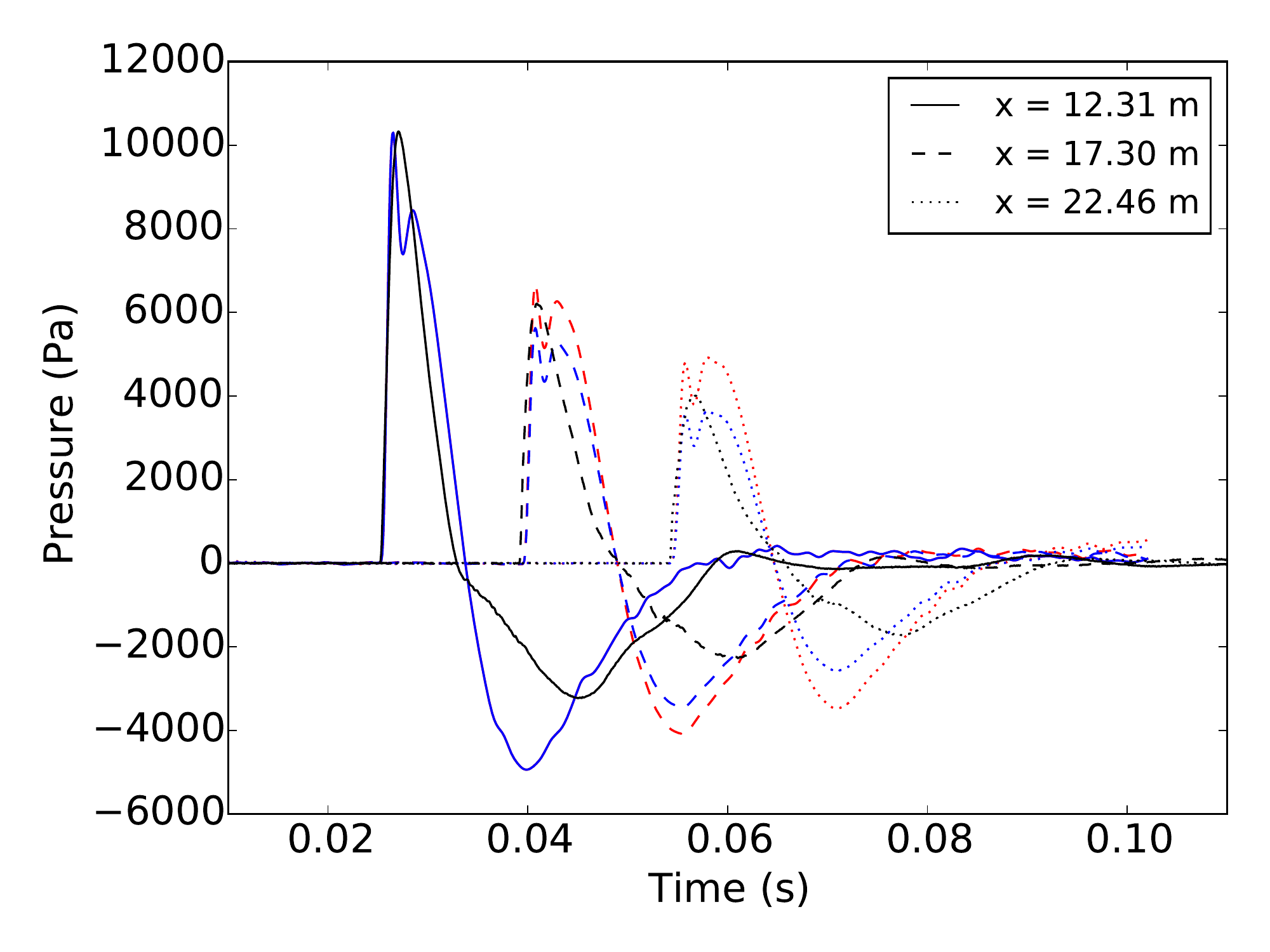}
\end{minipage} \\
\begin{minipage}[t]{2mm} 
\large b) 
\end{minipage}
\begin{minipage}[t]{85mm}
\vspace{-10pt}
\includegraphics[width=1\textwidth]{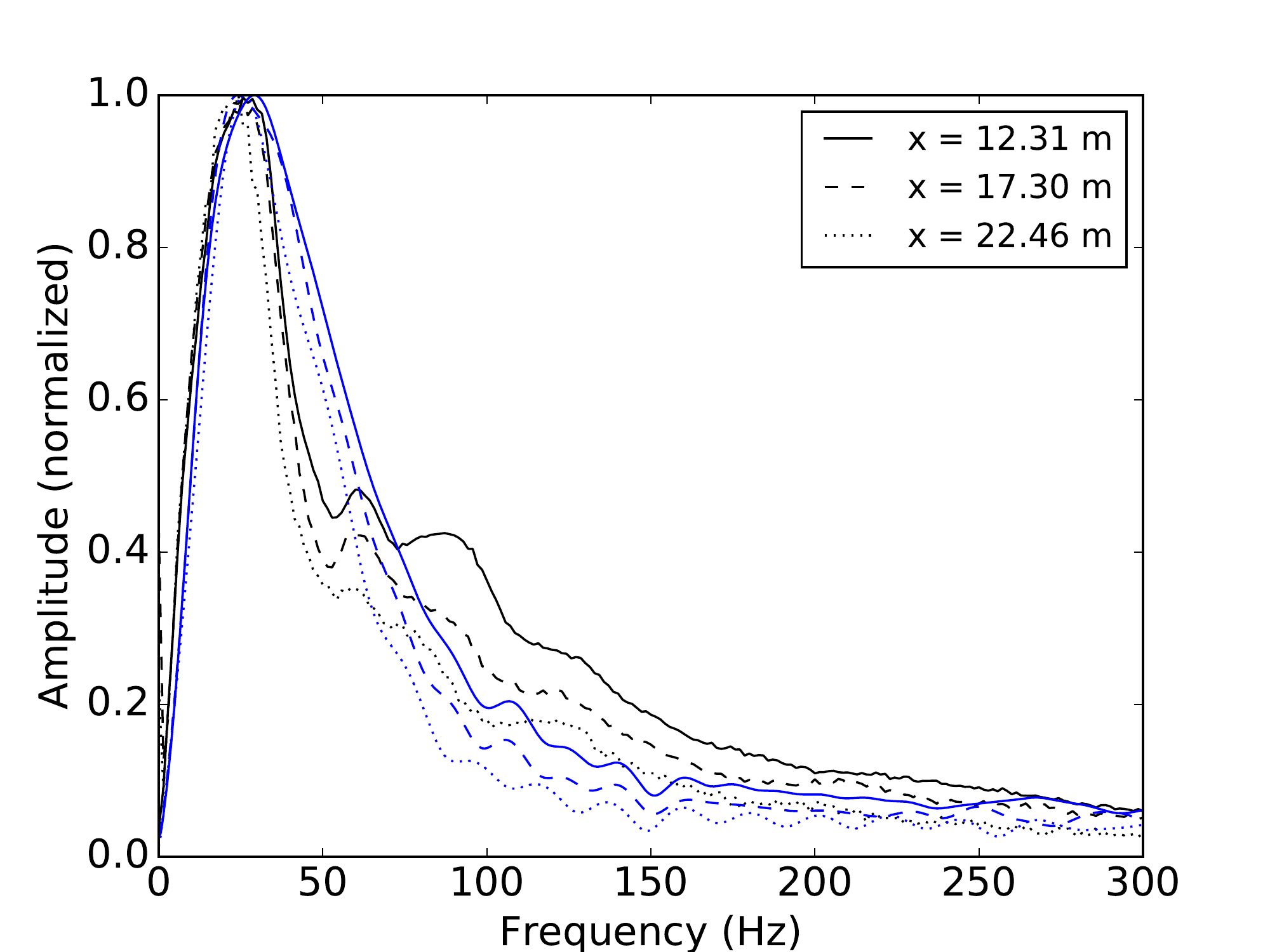}
\end{minipage} \\
\caption{\label{fig:air-pressure} 
Time series of the pressure a) and the normalized amplitude spectrum b) for the simulated (color) and measured (black) air pressure 5~cm above the snowpack. 
The red (cylindrical) and blue (spherical) curves show the relatively small effect of the spreading correction necessary because of the 2D equations in the simulation.
The velocity of the air pressure wave in the simulation is $V_{air} = 350$\,m/s.}
\end{figure}

The air pressure measurements above the snow surface and corresponding frequency spectra are shown in Figure \ref{fig:air-pressure} and compared to the corresponding results of the numerical simulation.
As the equations in the simulation are 2D a correction accounting for the differences between cylindrical and spherical spreading is applied for accuracy and completeness. 
In the 2D simulation the waves propagate cylindrically from the point of explosion and the amplitude decays proportional to $1/\sqrt{r}$, but in the field measurements the source is a point source which shows a spherical amplitude decay proportional to $1/r$ \citep{aki:1980}.
The differences resulting from this processing step are rather small because of the relatively large distance from the source which leads to a small curvature of the waves. The differences are shown in Figure \ref{fig:air-pressure}a). The simulated air pressure is shown in red while the air pressure corrected for spherical spreading is shown in blue. Omitting this step would not lead to any different conclusions for this study.

The length of a typical signal is around 40~ms and the main frequency content is between 20\,Hz and 70\,Hz. 
The two receivers at larger distances from the point of the explosion show a stronger decrease of frequencies between $\sim$ 50\,Hz and 150\,Hz than the one closest to the point of explosion.
Experiments with similar charge sizes and elevations of the point of explosion show very similar pressure wave forms and amplitudes.

\subsection{Acceleration in the snowpack}

\begin{figure}
\centering
\large a) 
\begin{minipage}[t]{85mm}
\vspace{-10pt}
\includegraphics[width=1\textwidth]{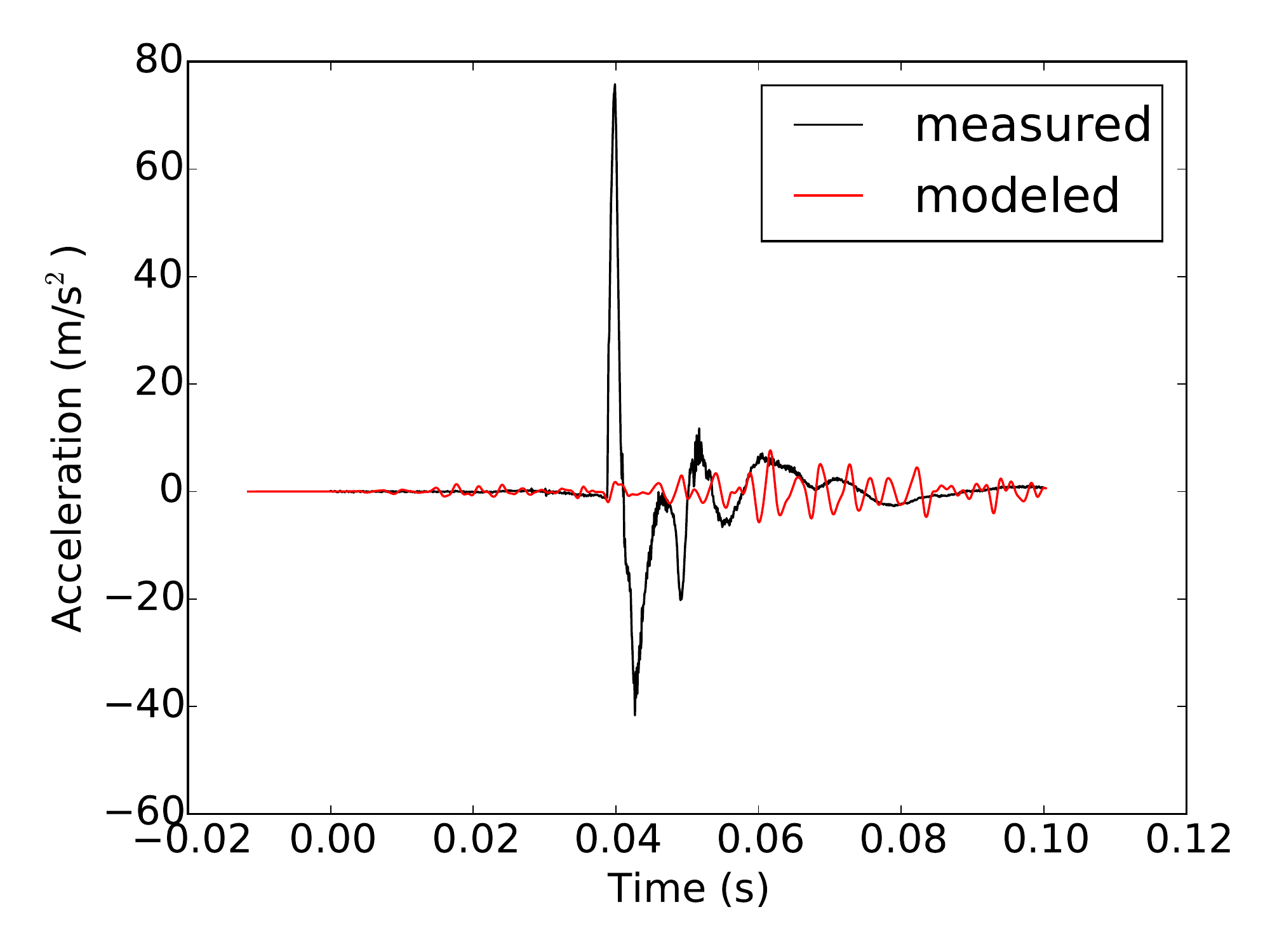}
\end{minipage} \\
\begin{minipage}[t]{2mm}
\large b) 
\end{minipage}
\begin{minipage}[t]{85mm}
\vspace{-10pt}
\includegraphics[width=1\textwidth]{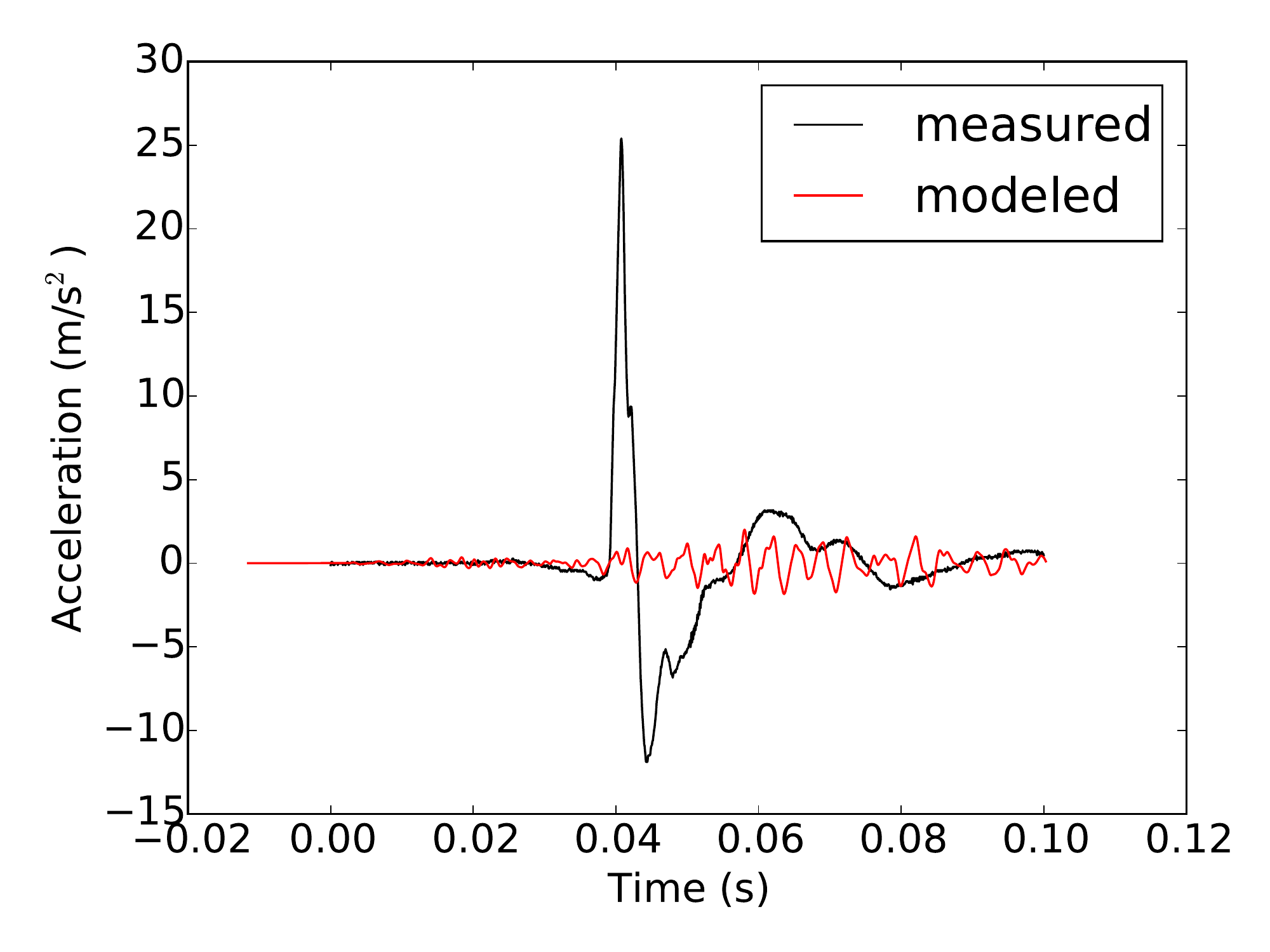}
\end{minipage} \\ 
\begin{minipage}[t]{2mm}
\large c) 
\end{minipage}
\begin{minipage}[t]{85mm}
\vspace{-10pt}
\includegraphics[width=1\textwidth]{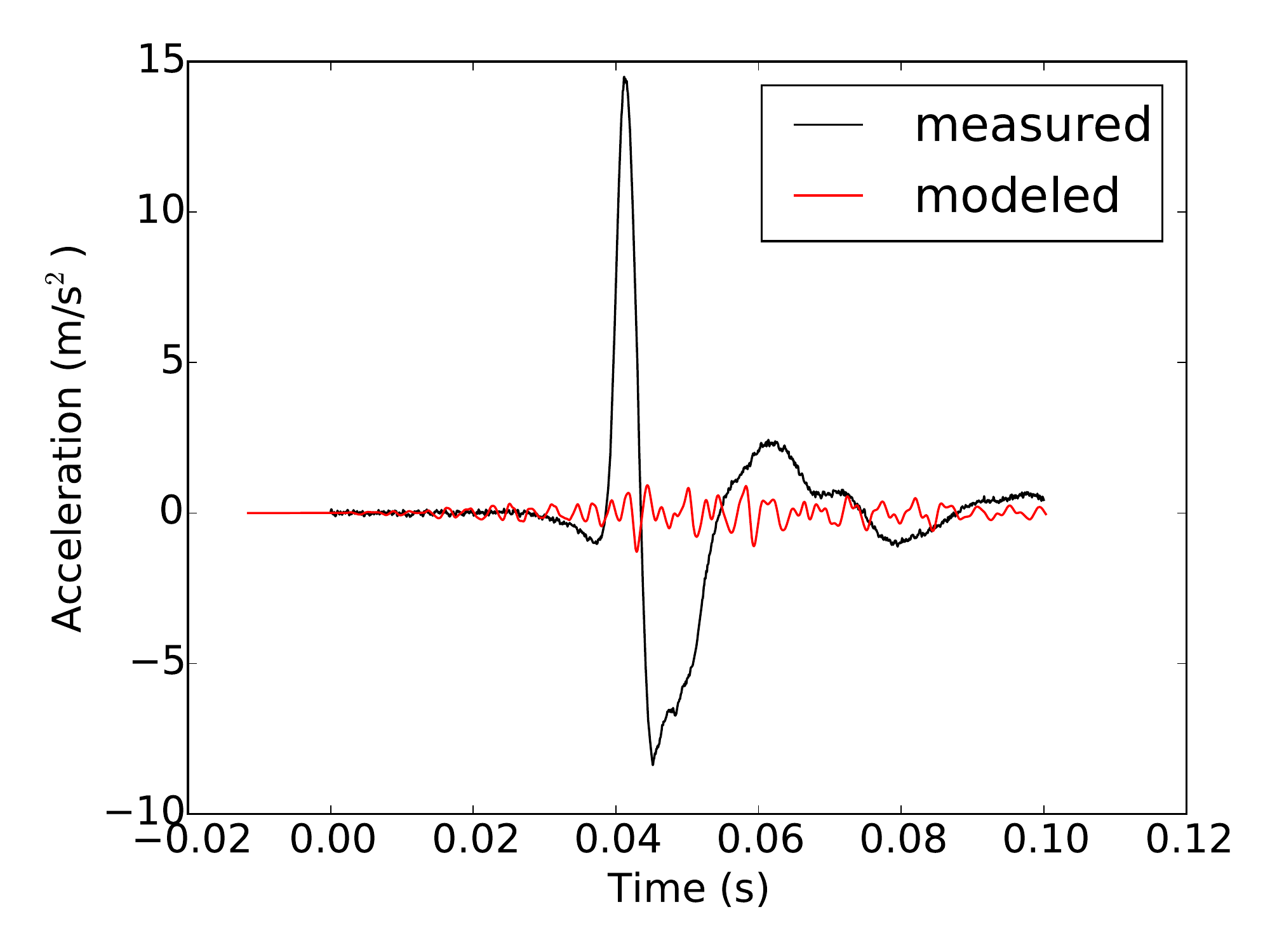}
\end{minipage}
\caption{\label{fig:17m-skeleton} Comparison of measured acceleration with modeled acceleration of the ice skeleton of the snow at a horizontal distance of 17.3\,m. The receivers were buried a) 0.13\,m, b) 0.48\,m, and c) 0.83~m blow the snow surface.}
\end{figure}

Simulated vertical acceleration of the ice skeleton 0.13~m, 0.48~m, and 0.83~m below the snow surface at a horizontal distance of 17.3~m from the explosion are compared to the measured acceleration at the same locations in Figure~\ref{fig:17m-skeleton}.
The measured accelerations show a strong peak of acceleration at about 39~ms. This is approximately the same time as the air pressure wave arrives at the pressure sensor in the air above the snowpack. However, before the strong peak smaller peaks and negative acceleration are present in the field recordings.
The simulated ice matrix acceleration shows small peaks of arriving wave fronts beginning approximately 15~ms after the triggering of the explosion. This earlier arrival corresponds to the fact that the acoustic wave speeds in the snowpack is higher than the wave speed in the air. 
Yet, the prominent recorded peak at 39~ms is considerably smaller in the simulated matrix acceleration. Subsequent arriving wave modes have a similar amplitude in the simulation and in the field measurements.  

\begin{figure}
\centering
\large a) 
\begin{minipage}[t]{85mm}
\vspace{-10pt}
\includegraphics[width=1\textwidth]{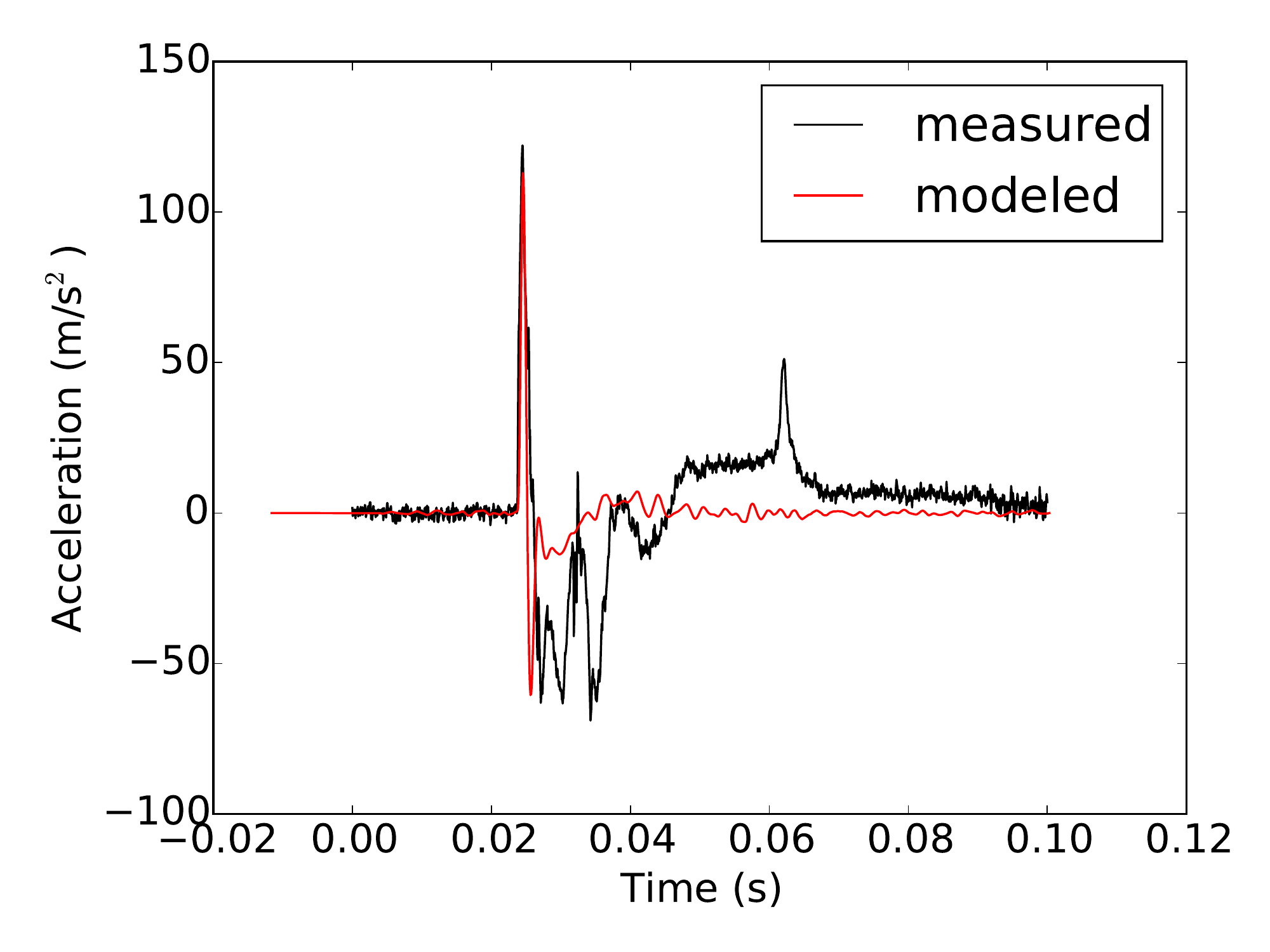}
\end{minipage} \\
\begin{minipage}[t]{2mm}
\large b) 
\end{minipage}
\begin{minipage}[t]{85mm}
\vspace{-10pt}
\includegraphics[width=1\textwidth]{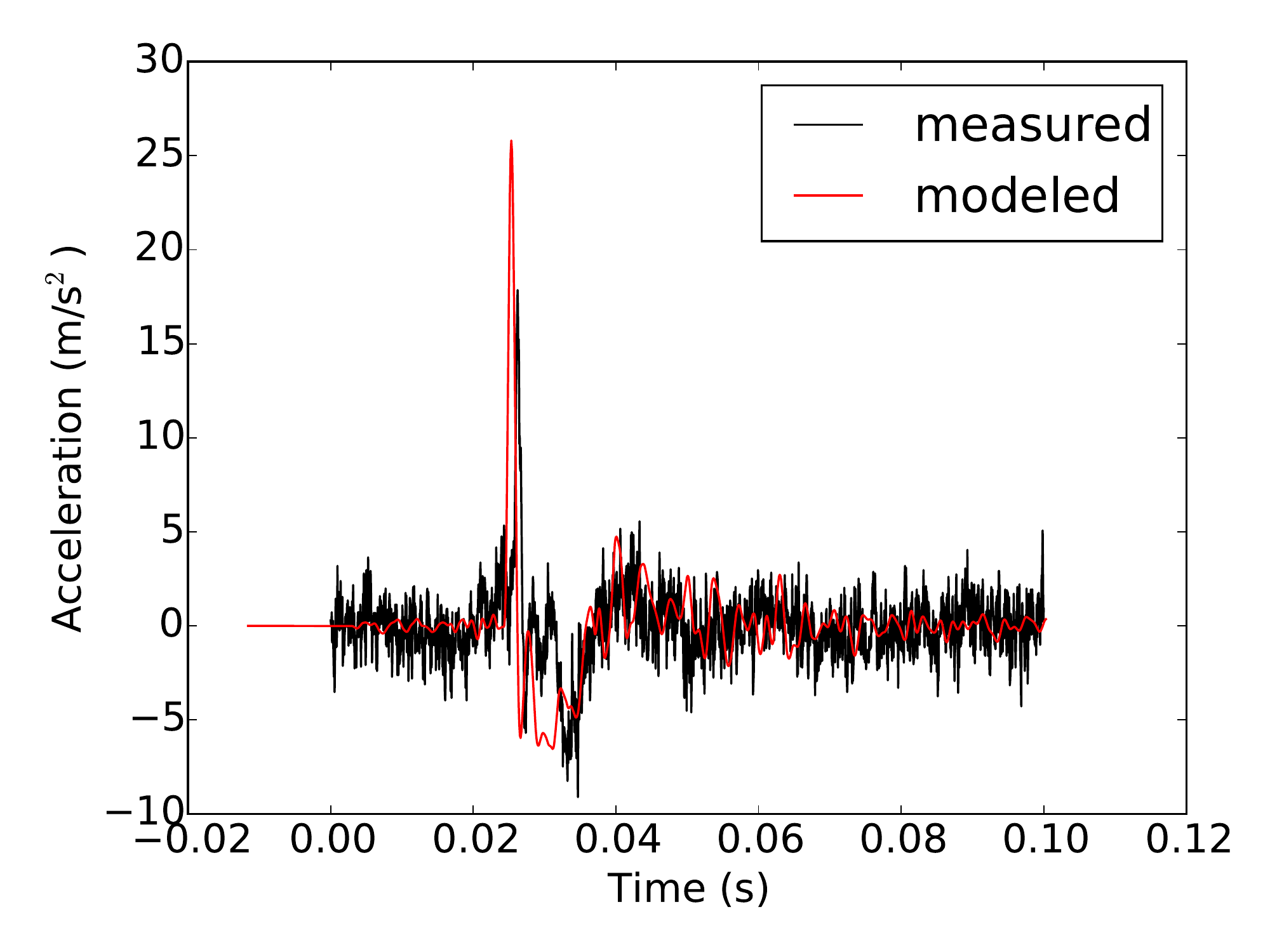}
\end{minipage} \\
\begin{minipage}[t]{2mm}
\large c) 
\end{minipage}
\begin{minipage}[t]{85mm}
\vspace{-10pt}
\includegraphics[width=1\textwidth]{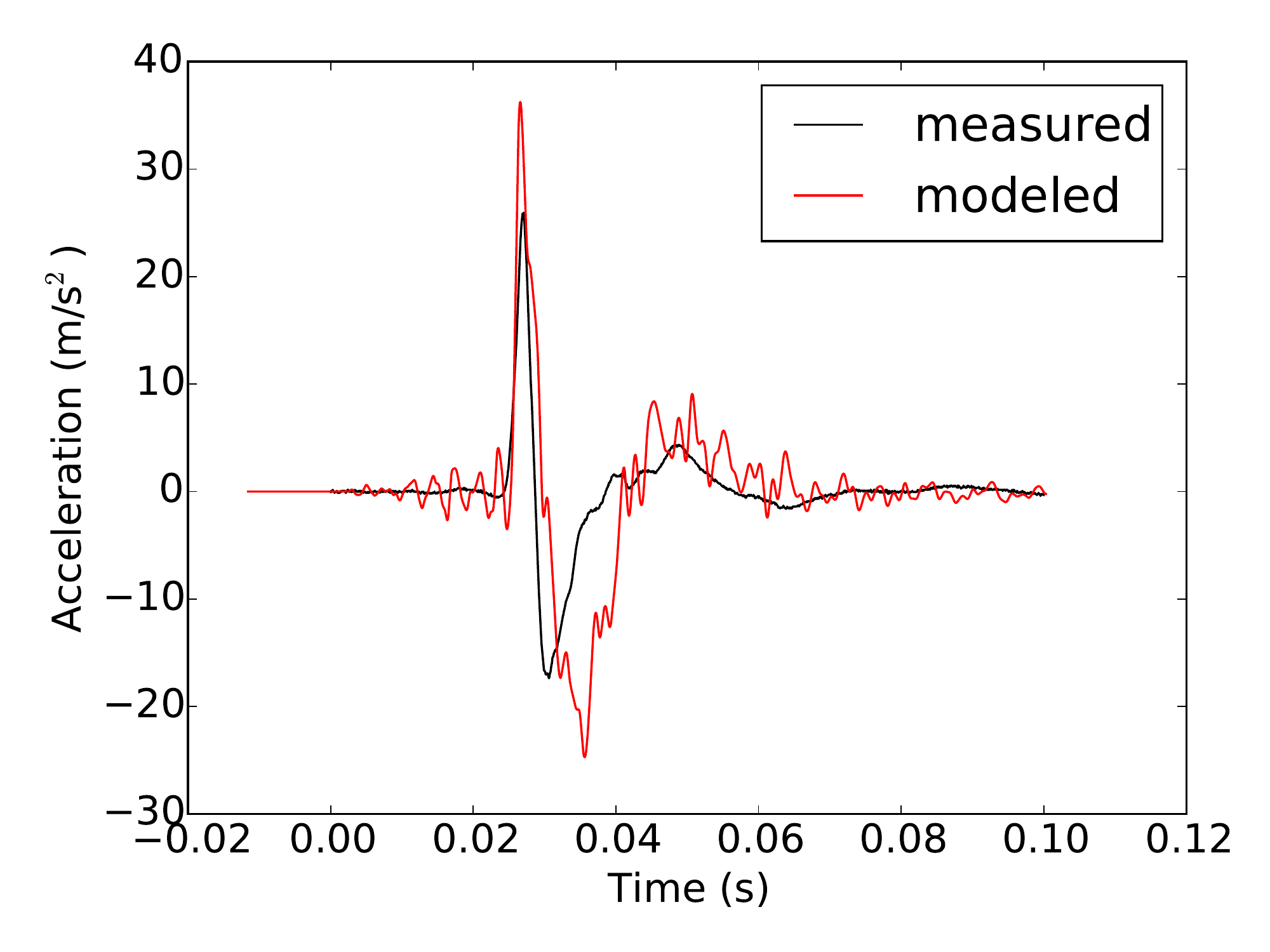}
\end{minipage} 
\caption{\label{fig:12m-pore} Measured acceleration compared to simulated acceleration of the pore fluid at a horizontal distance of 12.3\,m. The receivers were buried a) 0.13\,m, b) 0.48\,m, and c) 0.83~m blow the snow surface.}
\end{figure}

\begin{figure}
\centering
\large a) 
\begin{minipage}[t]{85mm}
\vspace{-10pt}
\includegraphics[width=1\textwidth]{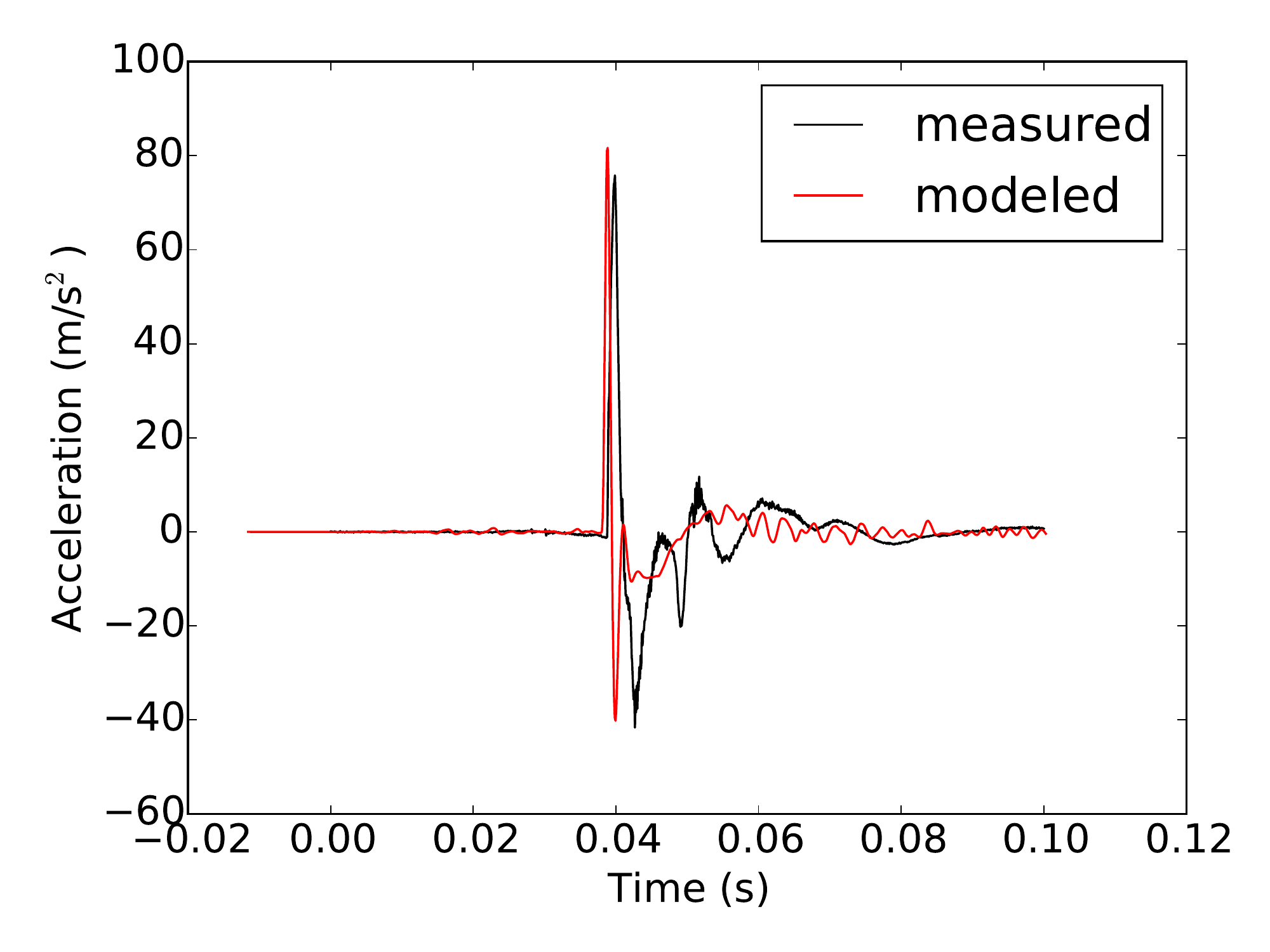}
\end{minipage} \\
\begin{minipage}[t]{2mm}
\large b) 
\end{minipage}
\begin{minipage}[t]{85mm}
\vspace{-10pt}
\includegraphics[width=1\textwidth]{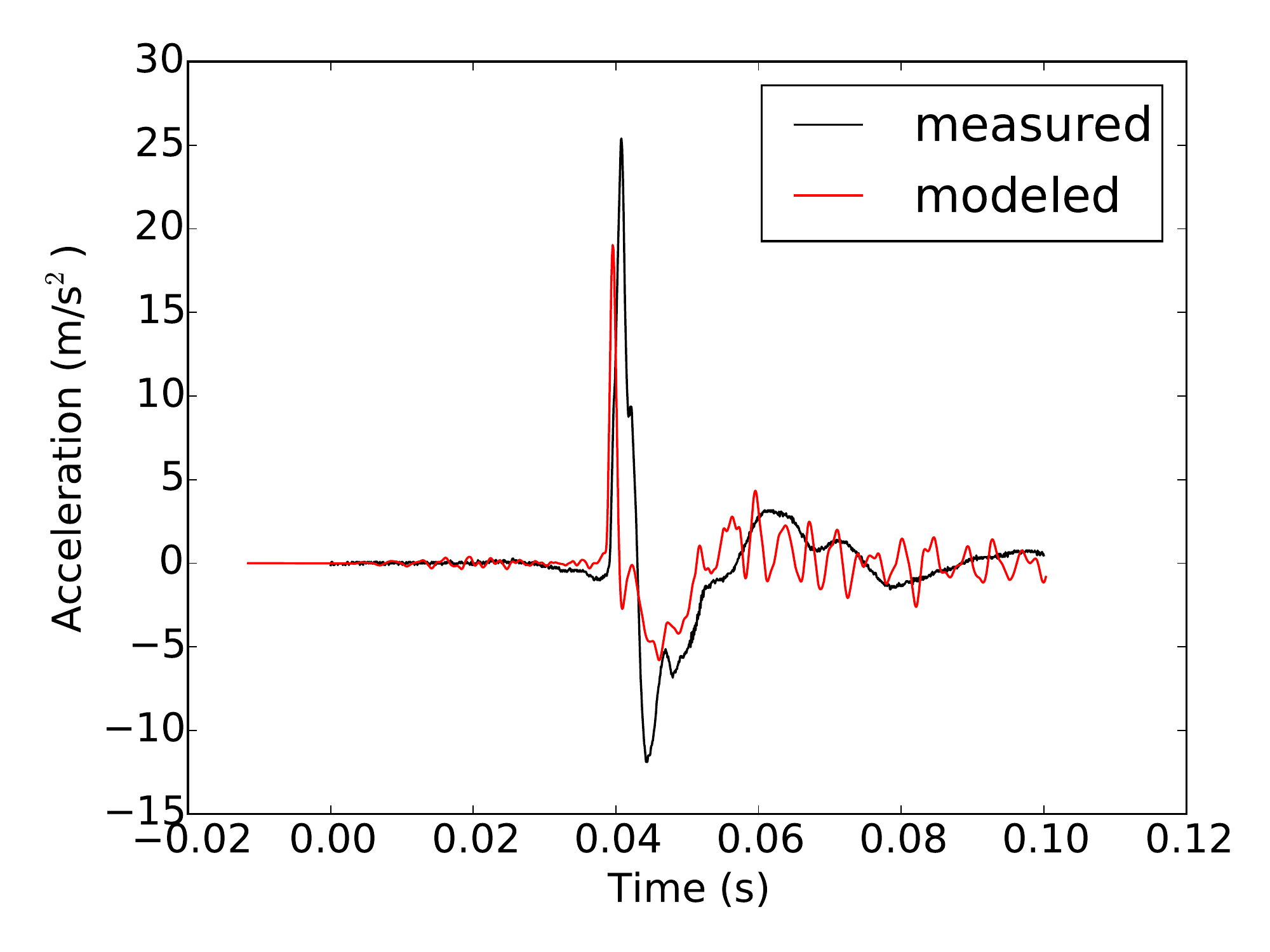}
\end{minipage} \\ 
\begin{minipage}[t]{2mm}
\large c) 
\end{minipage}
\begin{minipage}[t]{85mm}
\vspace{-10pt}
\includegraphics[width=1\textwidth]{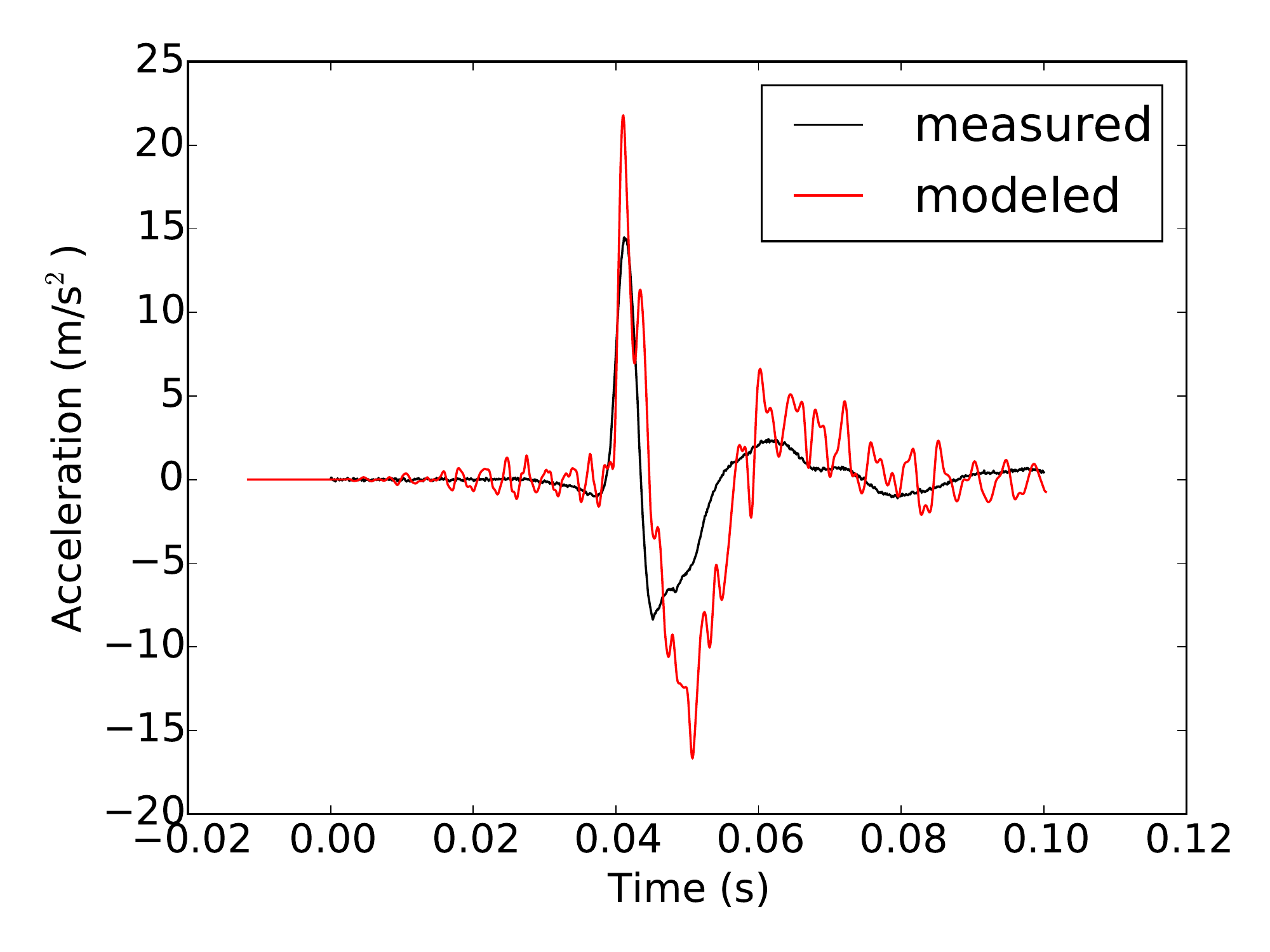}
\end{minipage}
\caption{\label{fig:17m-pore} Same as Figure \ref{fig:12m-pore}, but for a horizontal distance of 17.3\,m.}
\end{figure}

\begin{figure}
\centering
\large a) 
\begin{minipage}[t]{85mm}
\vspace{-10pt}
\includegraphics[width=1\textwidth]{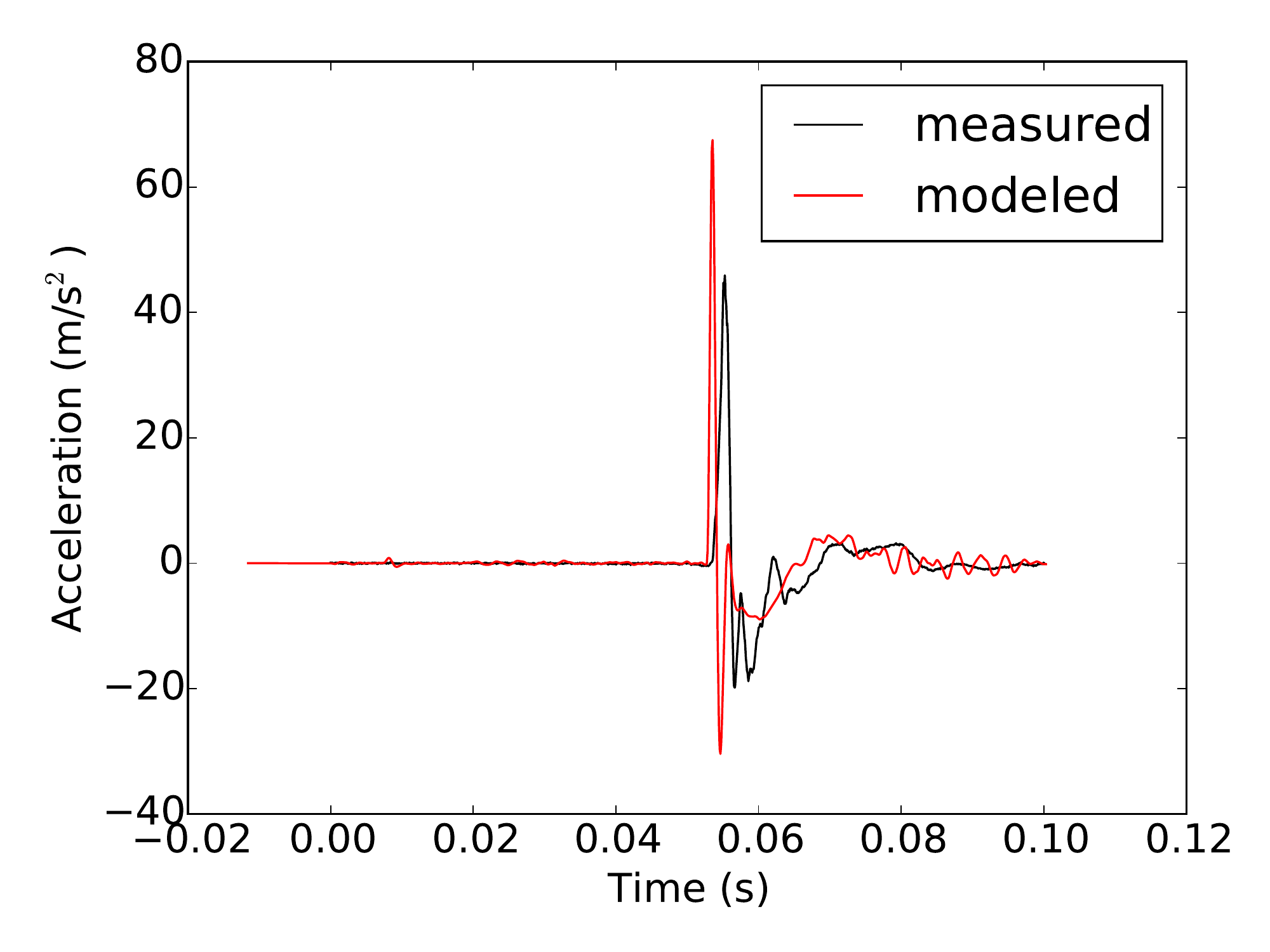}
\end{minipage} \\
\begin{minipage}[t]{2mm}
\large b) 
\end{minipage}
\begin{minipage}[t]{85mm}
\vspace{-10pt}
\includegraphics[width=1\textwidth]{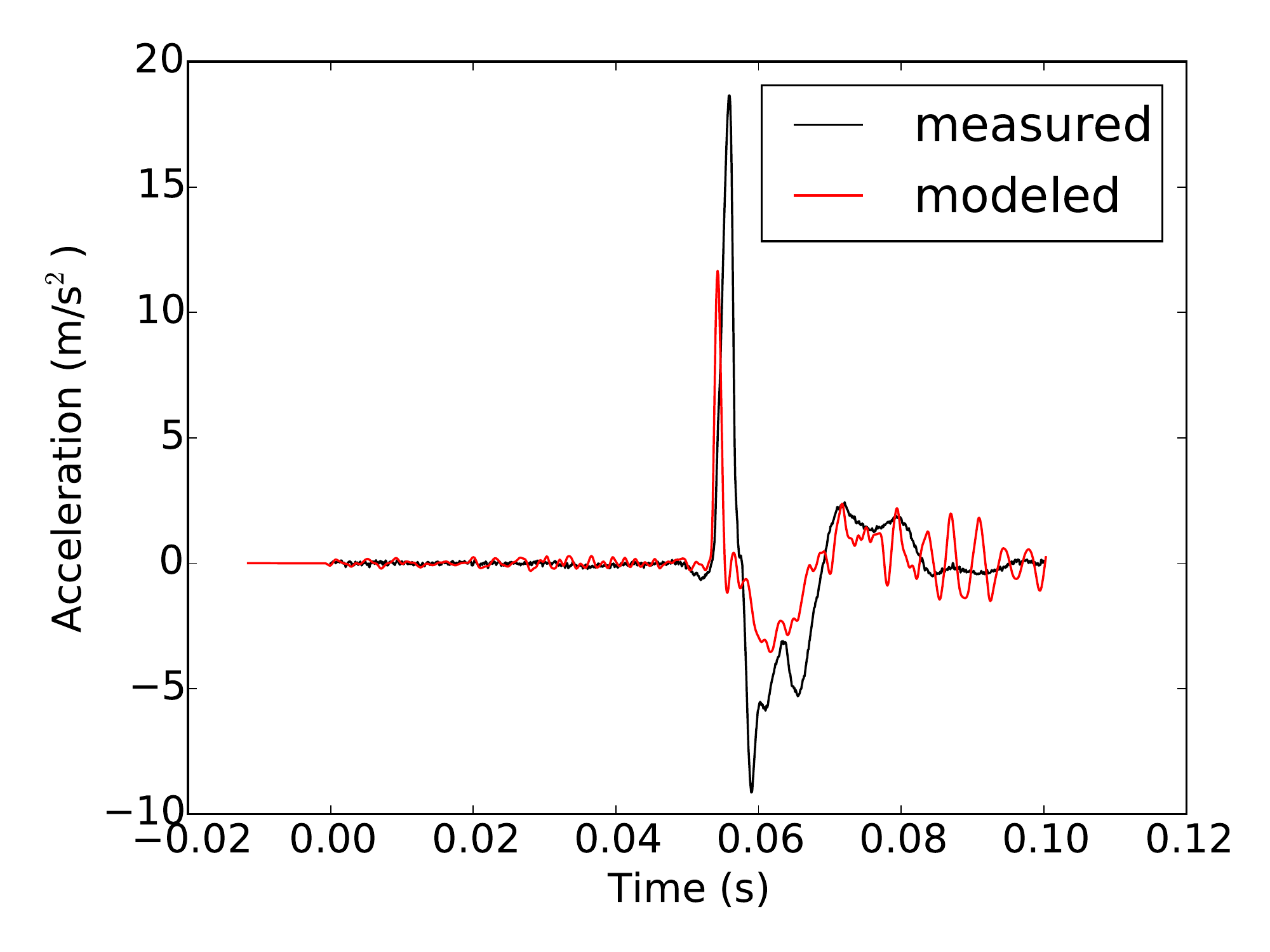}
\end{minipage} \\ 
\begin{minipage}[t]{2mm}
\large c) 
\end{minipage}
\begin{minipage}[t]{85mm}
\vspace{-10pt}
\includegraphics[width=1\textwidth]{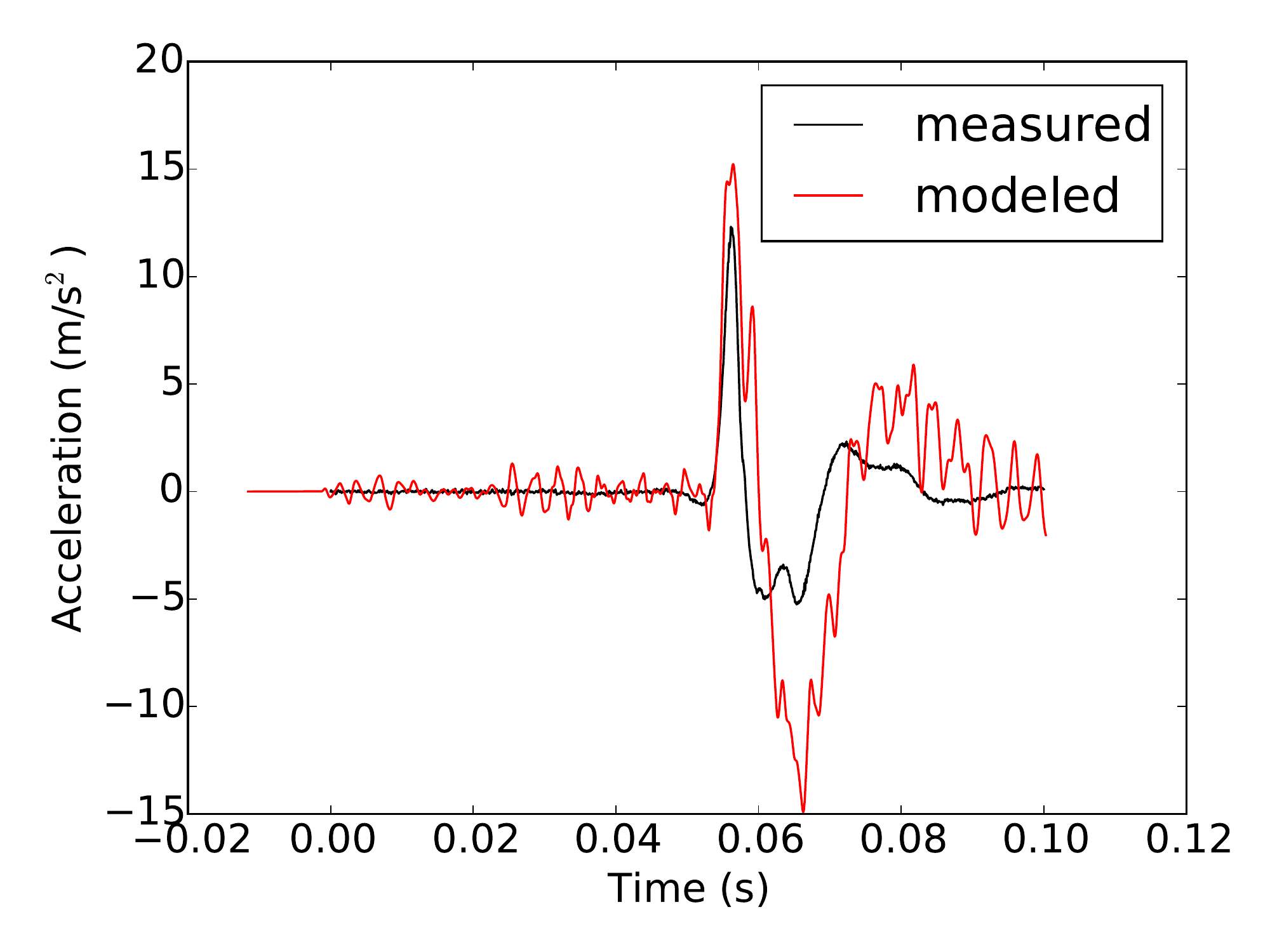}
\end{minipage} 
\caption{\label{fig:22m-pore} Same as Figure \ref{fig:12m-pore}, but for a horizontal distance of 22.4\,m.}
\end{figure}

The measured accelerations are compared to the pore fluid accelerations in Figures \ref{fig:12m-pore}, \ref{fig:17m-pore}, and \ref{fig:22m-pore} for horizontal distances of 12.3~m, 17.3~m, and 22.5~m from the point of the explosion, respectively.
Here, the prominent peak in the measured acceleration corresponds to the simulated air pressure wave propagating through the pore space of the snow pack. 
As the arrival times in the ice matrix responds well with the expected wave speeds in the snowpack we think that the response of the ice matrix is superimposed by the stronger response of the pore fluid.

From the recording time of the modeled seismograms 11.7~ms is subtracted to account for the higher air wave velocities due to non-linear effects in the vicinity of the explosion.
The numerically simulated accelerations shown in Figures~\ref{fig:17m-skeleton}, \ref{fig:12m-pore}, \ref{fig:17m-pore}, and \ref{fig:22m-pore} were scaled with a factor $10^{-2}$ (-40~dB).
This scaling factor reproduces the coupling effect well for the receivers located 0.13~m and 0.48~m below the snow surface.
For the receivers at a depth of 0.83~m below the snow surface the scaling factor is $10^{-1}$ (-20~dB).
A part of this lower amplitude of the measured seismograms compared to the simulated particle acceleration can be explained with the higher inertia of the receiver compared to an air particle. In the simulation the stresses at the receiver locations act on an air particle with a much lower mass than that of a physical receiver. The resulting acceleration from this stress will therefore be higher than the measured acceleration.
As stresses are applied to the receiver from both, the ice skeleton and the air in the pore space the response of the receiver will actually be a complex combination of the stress field, snow porosity, bulk modulus and density of the foam surrounding the receiver, and how well the foam is coupled to the ice skeleton.
A method to estimate the coupling of a receiver to a visco-elastic ocean bottom, which represents a somewhat similar situation, has been shown by \citet{sutton:1981}.

Vertical accelerations decrease rapidly with depth at all distances from the point of explosion. This effect can be explained with the strong attenuation for the second compressional wave.
For the receiver 0.83~m below the snow surface it is not exactly clear whether the lower scaling factor is due to an increased coupling between snow and receiver or if the lower amplitude of the simulation is due to the snowpack properties in the model. It is save to say that predictions for wave propagating in the pore fluid are better higher in the snowpack as less layers are involved. 
However, the most plausible explanation is better coupling to the ice matrix due to the lower porosity as an overestimated attenuation would also lead to strong dispersion for the second compressional wave \citep{johnson:1982, sidler:2015}.
Such a dispersion can not be seen in the receivers at a depth of 0.83~m below the snow surface.

Generally it can be observed that the wave field in the snowpack is complex and can not well be described by identifying individual wave modes. This is due to the interaction of the propagating waves with the many interfaces present in the simulation. On every interface waves are reflected, transmitted and converted into all supported wave modes. The proportions of the waves that get reflected, transmitted and converted depends on the material properties of of the involved snow layers and the incidence angles. Moreover if the wavelengths are longer than the involved layers the reflection, transmission and conversion originates from a combination of the snow layers `sensed' by the wave. In a direct numerical method where the wave field is evaluated on discrete grid points such as we use here these considerations are implicitly resolved by solving the underlying differential equations for the field variables. Here, these field variables are the horizontal and vertical matrix velocities, the relative horizontal and vertical pore fluid velocities, the horizontal and vertical stresses in the ice matrix, the shear stress in the ice matrix and the pore fluid pressure. 

As a consequence of the complex wave filed the shear and compressional stresses are difficult to observe in field measurements. In the simulation on the contrary all field variables can be evaluated individually and in combination with the field measurements it is possible to identify which field variables contribute to the recording.

\subsection{Snow failure}
\label{sec:strength}

\begin{figure}
\centering
\large a) 
\begin{minipage}[t]{85mm}
\vspace{-10pt}
\includegraphics[width=1\textwidth]{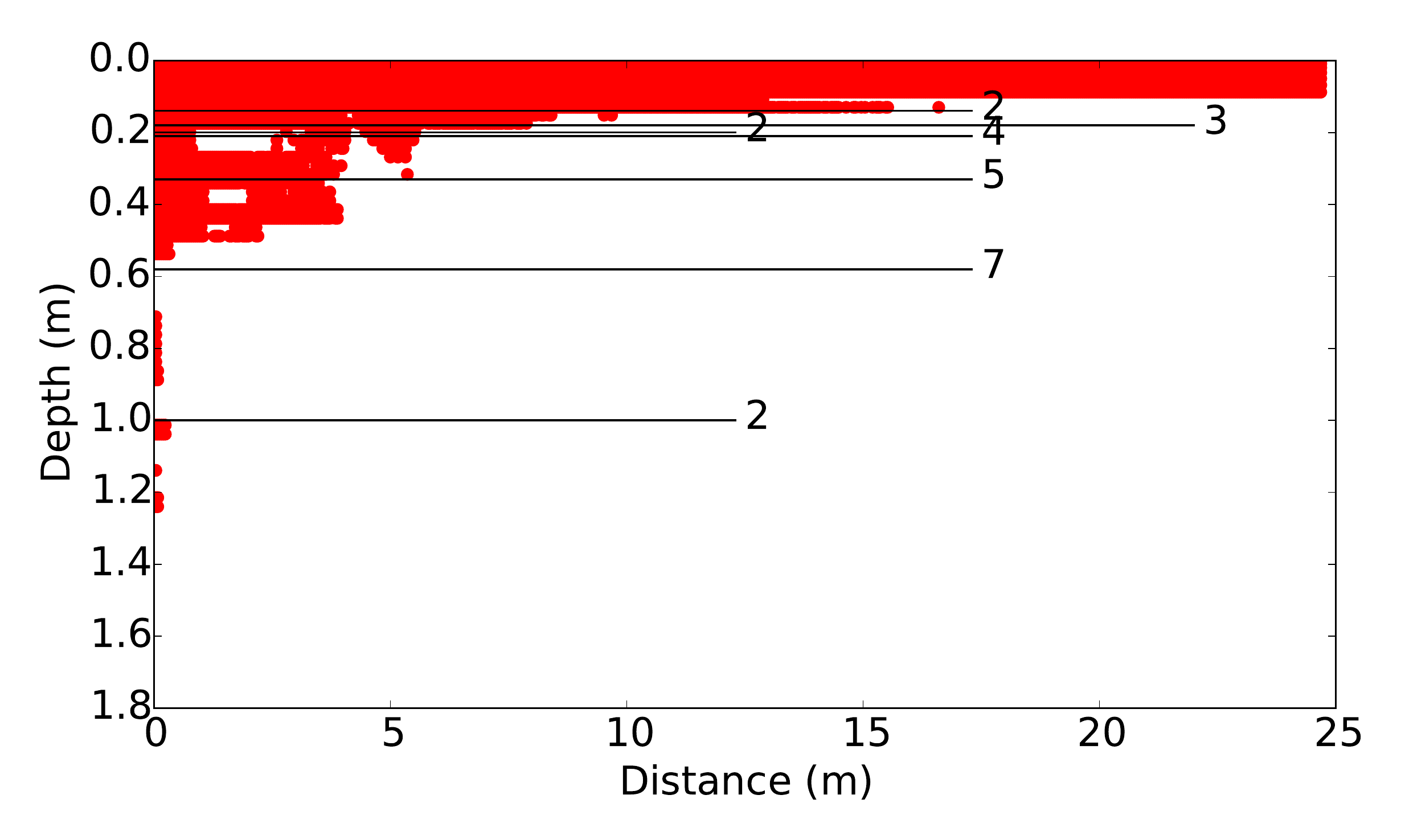}
\end{minipage} \\
\begin{minipage}[t]{2mm}
\large b) 
\end{minipage}
\begin{minipage}[t]{85mm}
\vspace{-10pt}
\includegraphics[width=1\textwidth]{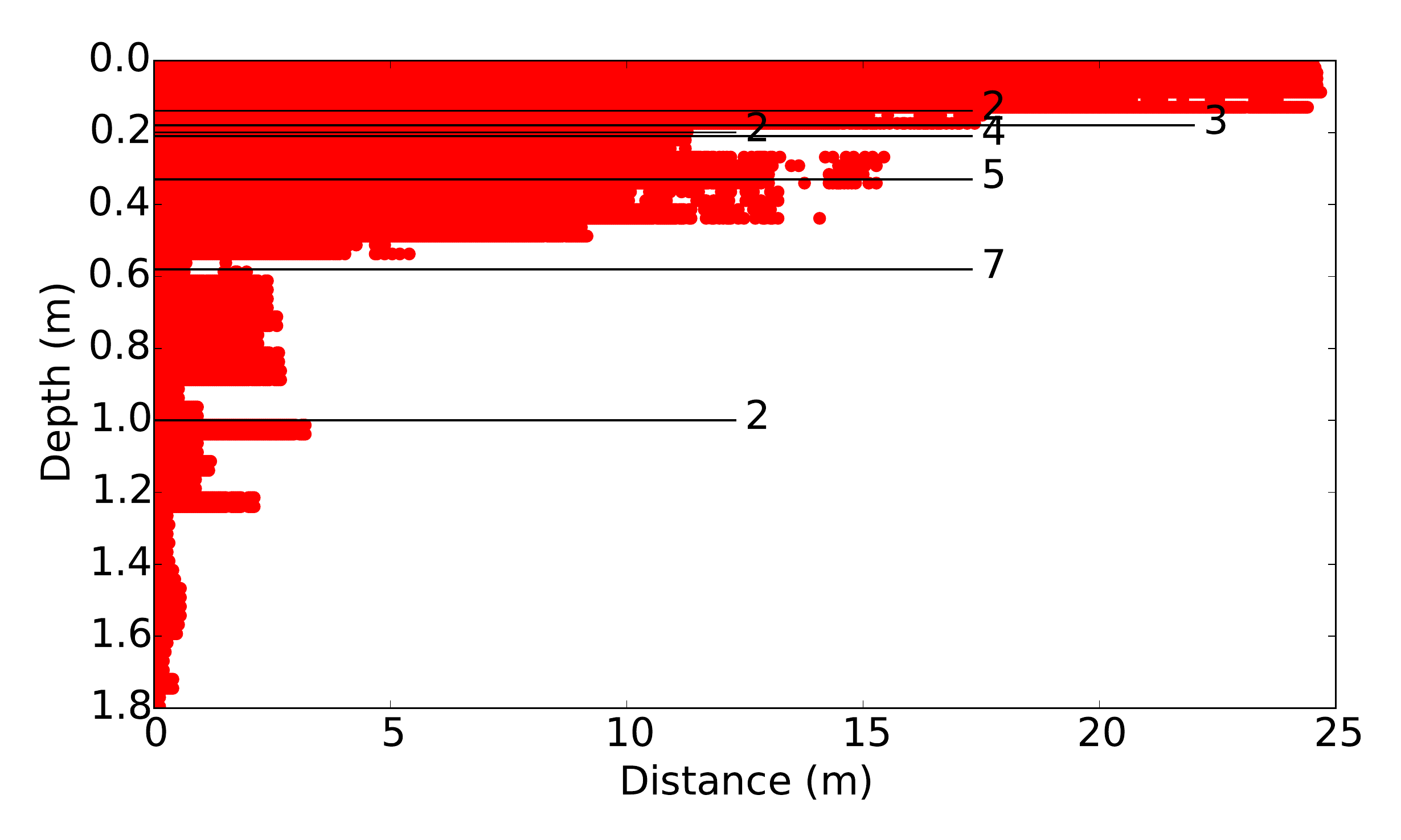}
\end{minipage} \\ 
\caption{\label{fig:failloc} Simulated locations of snow failure (red dots) due to principal a) normal and b) shear stress compared to observed failure locations (black lines). The number behind the black line indicates the number of the experiment after which the failure was observed. }
\end{figure}

On the investigated test day, eight experiments were performed. 
The charge sizes and charge elevation above the snow surface for the eight experiments are shown in Table \ref{tab:charges}. 
%
\begin{table}
\caption{Explosive charges for experiments on 27 February 2014.}
\begin{center}
\begin{tabular}{ccc}
\hline
\# of experiment & Charge size (kg) & Elevation above snow surface (m) \\
\hline
 	1	& 4.25	& 1.0  \\
	2	& 4.25	& 1.0  \\
 	3	& 4.25	& 1.9  \\
	4	& 4.25	& 2.0  \\
	5	& 5.0		& 2.0  \\
	6	& 5.0		& 3.0  \\
	7	& 5.0		& 3.0  \\
	8	& 10.0	& 2.0  \\
\hline
\end{tabular}
\end{center}
\label{tab:charges}
\end{table}%
Failure was observed mainly in the upper 30\,cm of the snowpack and at the weak layer at at a depth of about $\sim$1~m below the snow surface. 
Failures even deeper in the snowpack occurred with either very large charges or close to the point of the explosion. For some experiments, failure could not be observed at the close pit locations but only in the furthest pit. We assume that there was failure in the closer pits too, but could not be identified by analyzing the recorded video images.
Figure \ref{fig:failloc} shows the locations of the observed and simulated snow failure.
Observed failed layers are indicated in the figure with black horizontal lines. The number at the right end of line indicates the number of the experiment in which the layer failed.
In the field experiments the snow failure locations were only evaluated in the snow pits. If a snow pit showed layer failure at a specific depth it was assumed that this layer had also failed closer to the point of the explosion.
The points of failure obtained from the numerical simulation are indicated by red dots. Each red dot represents a grid point where the principal stress has exceeded the computed strength of the snowpack in one or more snapshots of the simulation. 

\begin{figure}
\centering
\includegraphics[width=1\textwidth]{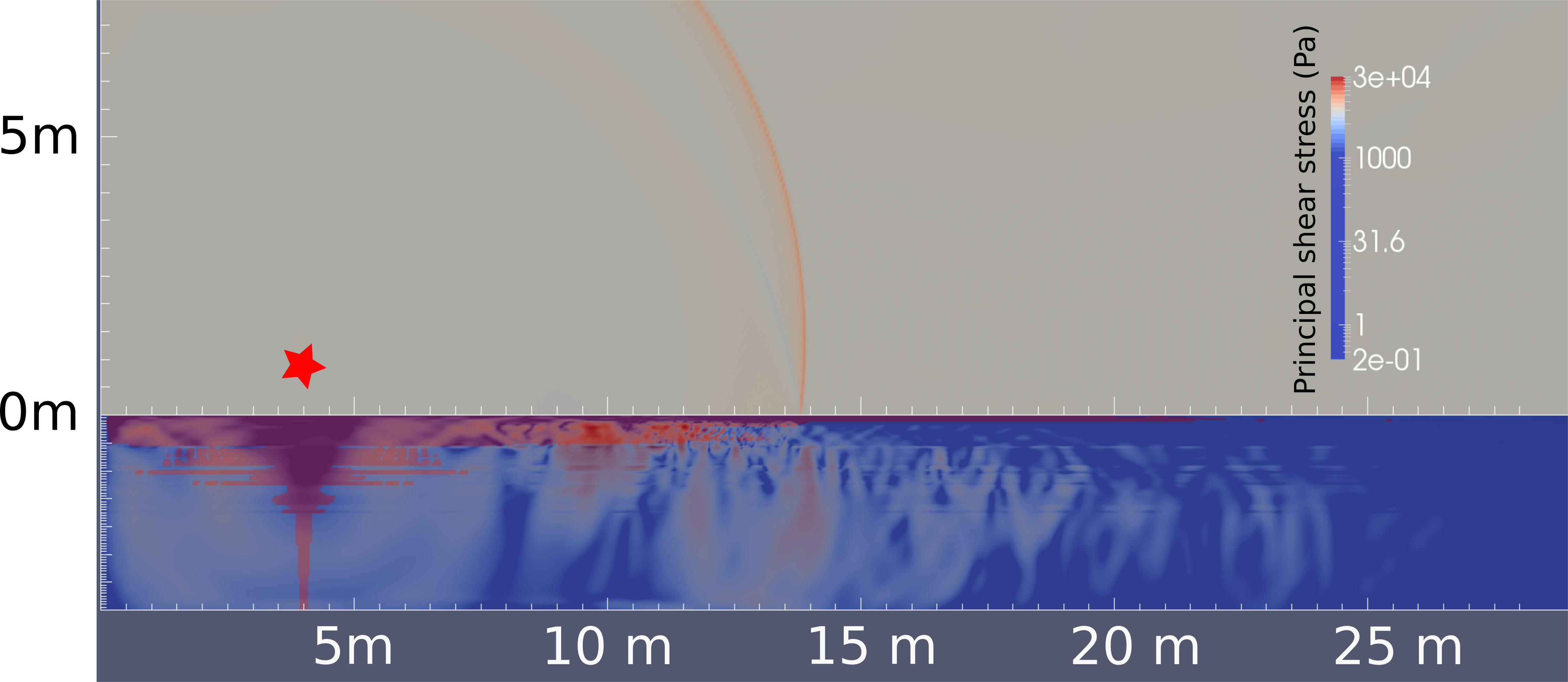}
\caption{\label{fig:snapshots} Snapshot after 30 ms of the numerical simulation. The red star marks the point of the source. Blue and white color indicates principal shear stress in the snowpack. Locations where the principal shear stress of the propagating waves has exceeded the shear strength of the snowpack in the simulation are marked with red color.}
\end{figure}

One of the evaluated snapshots after 30~ms simulation time is shown in Figure \ref{fig:snapshots}. In the upper acoustic domain, the air pressure wave can be seen as a red front at 14.5~m from the left boundary of the simulation. Principal shear stresses are shown in the lower poroelastic domain with blue and white color.
Locations where the principal shear stress exceeded the shear strength of the snowpack during the past 30~ms of the simulation are indicated with dark red color in the lower domain.

\section{Discussion}

\subsection{Air pressure above the snow surface}

The air pressure measurements of the simulation fit the measurements of the experiments if the amplitude is corrected for the geometrical spreading of a point source. Such a correction is straightforward for the pressure recordings, where the wave travels in a relatively straight path through a homogeneous medium.
Such a correction is more difficult for the recordings in the snowpack, where the waves are reflected and follow a complicated path between snow layers. Therefore, the measurements in the snowpack are not corrected for geometrical spreading and it can be assumed that the simulated wave field is slightly overestimating the amplitudes in the snowpack.

The time delay for the arrival of the air wave in the simulation and the higher than expected velocity for the speed of sound are presumably due to non-linear effects in the propagation of pressure waves originating from explosions \citep{simioni:2015}.
Close to an explosion the particle displacement is large enough that stresses are not linearly related to the strain as in Hook's law. In air, large strains yield faster propagating waves than smaller strains and the waveform tends to acquire a steep front. The overpressure of the explosion propagates faster than the speed of sound in the form of a shockwave.
The numerical simulation of the experiment does not take into account these non linear effects. 
However, at larger distances, where the particle displacement becomes smaller due to geometrical spreading, internal friction, and heat dissipation, the pressure wave caused by the explosion propagates in an elastic fashion.
The change of waveform due to the nonlinear effects can be taken into account by adequately choosing the shape of the source waveform \citep[e.g.,][]{vandererden:2005}. For example, as we did here, by choosing a Friedlander source wavelet.

The speed of sound, derived from the wave arrival time at the microphones and their distance, is around 350~m/s for the field experiment, which is higher than the expected 330~m/s for this temperature \citep{simioni:2015}.
The theoretical shockwave speed $v_{shock}$ for the measured overpressure at 12.3~m from the explosion can be caculated as
\begin{equation}
v_{shock} = \upsilon \sqrt{\frac{p_2 - p_1}{\upsilon_1 - \upsilon_2}}, 
\end{equation}
where $p$ is the pressure and $\upsilon = 1/ \rho$ is the specific volume \citep[][p. 363]{jaeger:2007}. The subscripts 1 and 2 correspond to the regions in front of and behind the shockwave, respectively.
For air, the ideal gas law can be used to obtain the density $\rho$ from the air pressure and temperature as
\begin{equation}
\rho = \frac{p}{R_{spec} T}, 
\end{equation}
where $T$ is the absolute temperature in Kelvin and $R_{spec} = 287.058$~J~kg$^{-1}$K$^{-1}$ is the 
specific gas constant for dry air.
The measured maximum air pressure at the microphone was $\sim$10~kPa in excess of the atmospheric pressure of $p_1 \cong 84$ kPa at the elevation of the study site.
This pressure difference alone does not explain a shockwave that is propagating faster than the speed of sound.
The observed velocity can be explained also with an increase of air temperature behind the wave front.
To obtain the observed speed of $\sim$350~m/s the air temperature behind the shockwave has to be 9~$^\circ$C higher than in front of the shockwave.

\subsection{Acceleration in the snowpack}

The measured acceleration corresponds much better with the simulated acceleration of the pore fluid than with the simulated acceleration of the ice matrix.
This is a rather surprising result as intuitively one would expect that the accelerometers are coupled mainly to the ice skeleton of the snow. 
However, given that the porosity of the snow is usually higher than 0.5 the volume around the receiver mainly consists of air. It makes therefore sense to assume that the motion of the air around the receiver is represented in the recordings.

The introduction of non physically based scaling factors is admittedly a flaw in the modeling process.
However, the fact that a single scaling factor can reproduce the amplitude of most of the acceleration recordings so well is a strong indication that there is indeed some kind of coupling process involved between the snow and the accelerometers. 

The acceleration of the pore fluid in the simulation is almost an order of magnitude higher than the acceleration of the skeleton. 
Due to the open pore boundary conditions and the high porosity at the snow surface the air pressure from the blast wave is transmitted mostly to the pore space and only to a lesser extent to the ice skeleton of the snow.
This wave propagating in the pore space of a porous material is sometimes also called the 'slow' wave which is of high interest in hydrogeophysics and hydro carbon exploration as it directly connects hydrological properties of the porous materials to seismic wave propagation. Unlike in snow, the slow wave is a diffusive wave mode in these environments and cannot be directly measured.

It is not clear why the simulated accelerations are higher than the measured accelerations.
There are several mechanism that can potentially contribute to such a discrepancy.
Complete coupling between the snow and the accelerometers may not be provided. 
Also, the coupling of the accelerometers is complicated due to the porous nature of the snowpack. The receiver is not only embedded in a solid material but in a combination of a fluid and a solid material. Due to the high porosity of the snow the receiver is actually mainly surrounded by air. The stiffer ice frame will, however, have a better coupling to the receiver than the pore fluid.
Due to differences of the density between the air in the snowpack and the receiver the stresses applied to the receivers lead to accelerations that are different from the accelerations that would arise when the stresses would be applied to snow.

\subsection{Snow failure}

The simulated failure locations shown in Figure \ref{fig:failloc} suggest that snow failure is caused in shear rather than compression or tension.
This finding may be one of the reasons why explosives triggered above the snow surface are more effective in triggering avalanches than charges that are deployed within the snowpack. 
Explosive charges produce almost exclusively compressional waves. Yet, when the charge is placed above the snow surface the compressional waves are converted at the snow surface into all existing wave modes which are, in this case, the first and second compressional waves, the shear wave, a reflected wave in the air above the snowpack and a surface wave within the snowpack. 

The snow model features a relatively sharp increase in density at a depth of $\sim$0.5~m. Wave amplitudes and snow failure below this interface are considerably smaller (Figure \ref{fig:failloc} and Fig. \ref{fig:snapshots}) due to the large difference in impedance and the resulting reflectance of incident waves. 
The same observation is also true for the field measurements with the exception of a failure at 0.6~m and 1~m below the snow surface.
Although the layer at 0.6~m only failed with a considerably larger charge and at 1~m depth a weak layer existed that might have favored crack propagation from a failure location closer to the point of explosion.

This study is in many aspects complementary to the study of \citet{miller:2011}. While here we are missing the non-linear effects, the explicit source characterization, and the deformation of the snowpack in the immediate vicinity of the explosion we account for the porous nature of the snowpack and linear wave propagation at distances up to several tens of meters.
The comparison of the simulated with the measured accelerations shown in this study reveal the importance of the porous nature of the snowpack. Most of the recorded signal coincides with the particle movement in the air and not with the ice skeleton of the snowpack.
The porosity of the snowpack not only affects wave propagation inside the snowpack but also reflection and transmission of energy at the interface between the snowpack and the air above the snowpack.
In a highly porous material as snow a large fraction of the transmitted energy is converted into the second compressional wave that is propagating in the air of the snowpack and is strongly attenuated.
Neglecting the porous nature of the interface will consequently lead to overestimating of the energy propagating in the ice skeleton of the snow.

\section{Conclusions}

We have compared field measurements of air pressure and accelerations within the snowpack caused by the detonation of an explosive charge over a horizontal snowpack to a numerical simulation of an acoustic source over a Biot-type porous material.
The properties of snow as a porous material were derived from density profiles by using {\it a priori} information and empirical relationships.
The interface between the air and the porous material in the simulation was considered to be of the open-pore type.
The non-linear effects in the vicinity of the explosion were considered by using a Friedlander wave form for the source that was calibrated by the closest measurement of air pressure above the snowpack.

The field measurements consisted of air pressure measurements 0.05~m above the snow surface and acceleration measurements in the snowpack. Accelerometers were installed at three snowpits to measure snowpack accelerations. 
In each snow pit the acceleration was measured at depths of 0.13~m, 0.48~m, and 0.83~m below the snow surface.
Regularly spaced markers were placed on the pit walls in the three pits and were monitored with video cameras during the experiments. From the video images the failure locations were determined by visual inspection. 

The best fit of the amplitudes of the air pressure measurements was obtained when the air pressure in the simulation was corrected for the spherical spreading of a point source.
The acceleration recordings in the snowpack fitted the modelled acceleration of the air moving inside the snowpack well. The simulated ice accelerations are missing the characteristic peak acceleration that is specific to the measured accelerations.
This finding suggests that waves propagating the pore space, i.e. in the air, significantly contribute to wave propagation in snow due to an explosion.
Such waves can be simulated by porous models only and are not considered in the standard elastic or viscoelastic seismological models.

Snow failure locations in the simulation were evaluated by comparing the principal normal and shear stresses to snow strength.
Snow strength was considered to be a fraction of the bulk or shear modulus of the snow and was derived from the porosity model of the simulation. Snapshots of the principal normal and shear stresses for every 0.4~ms of the simulation were then evaluated for locations were the stresses exceeded the local strength of the snow.
The simulation results suggest that observed failures were mainly due to loading by shear stress.

\section*{Acknowledgements}
Rolf Sidler was funded by a fellowship of the Swiss National Science Foundation.
This study was partly funded by the Swiss Federal Office for the Environment FOEN.

\bibliographystyle{model2-names} 
\bibliography{0-references}

\begin{thebibliography}{40}
\expandafter\ifx\csname natexlab\endcsname\relax\def\natexlab#1{#1}\fi
\expandafter\ifx\csname url\endcsname\relax
  \def\url#1{\texttt{#1}}\fi
\expandafter\ifx\csname urlprefix\endcsname\relax\def\urlprefix{URL }\fi
\providecommand{\eprint}[2][]{\url{#2}}
\providecommand{\bibinfo}[2]{#2}
\ifx\xfnm\relax \def\xfnm[#1]{\unskip,\space#1}\fi
\bibitem[{Aki and Richards(1980)}]{aki:1980}
\bibinfo{author}{Aki, K.}, \bibinfo{author}{Richards, P.G.},
  \bibinfo{year}{1980}.
\newblock \bibinfo{title}{Quantitative seismology}.
\newblock \bibinfo{publisher}{W. H. Freeman}, \bibinfo{address}{New York}.
\bibitem[{Albert et~al.(2013)Albert, Taherzadeh, Attenborough, Boulanger and
  Decato}]{albert:2013}
\bibinfo{author}{Albert, D.G.}, \bibinfo{author}{Taherzadeh, S.},
  \bibinfo{author}{Attenborough, K.}, \bibinfo{author}{Boulanger, P.},
  \bibinfo{author}{Decato, S.N.}, \bibinfo{year}{2013}.
\newblock \bibinfo{title}{Ground vibrations produced by surface and
  near-surface explosions}.
\newblock \bibinfo{journal}{Applied Acoustics} \bibinfo{volume}{74},
  \bibinfo{pages}{1279--1296}.
\bibitem[{Binger and Miller(2015)}]{binger:2015}
\bibinfo{author}{Binger, J.B.}, \bibinfo{author}{Miller, D.A.},
  \bibinfo{year}{2015}.
\newblock \bibinfo{title}{Soft and hard slab snow dynamic response to
  explosives used in avalanche hazard mitigation}.
\newblock \bibinfo{journal}{Journal of Cold Regions Engineering} ,
  \bibinfo{pages}{04015003}.
\bibitem[{Biot(1956)}]{biot:1956}
\bibinfo{author}{Biot, M.A.}, \bibinfo{year}{1956}.
\newblock \bibinfo{title}{Theory of propagation of elastic waves in a
  fluid-saturated porous solid. {I}. {L}ow-frequency range}.
\newblock \bibinfo{journal}{Journal of the Acoustical Society of America}
  \bibinfo{volume}{28}, \bibinfo{pages}{168--178}.
\bibitem[{Biot(1962)}]{biot:1962}
\bibinfo{author}{Biot, M.A.}, \bibinfo{year}{1962}.
\newblock \bibinfo{title}{Mechanics of deformation and acoustic propagation in
  porous media}.
\newblock \bibinfo{journal}{Journal of Applied Physics} \bibinfo{volume}{33},
  \bibinfo{pages}{1482--1498}.
\bibitem[{Bones et~al.(2012)Bones, Miller and Savage}]{bones:2012}
\bibinfo{author}{Bones, J.}, \bibinfo{author}{Miller, D.},
  \bibinfo{author}{Savage, S.}, \bibinfo{year}{2012}.
\newblock \bibinfo{title}{An experimental dynamic response study of hard slab
  seasonal snow to explosive control}, in: \bibinfo{booktitle}{International
  Snow Science Workshop ISSW 2012, Anchorage AK, U.S.A., 16-21 September 2012},
  pp. \bibinfo{pages}{142--148}.
\bibitem[{Boyd(2001)}]{boyd:2001}
\bibinfo{author}{Boyd, J.P.}, \bibinfo{year}{2001}.
\newblock \bibinfo{title}{Chebyshev and Fourier spectral methods}.
\newblock \bibinfo{publisher}{Dover Publications}, \bibinfo{address}{New York}.
  \bibinfo{edition}{2nd} edition.
\bibitem[{Carcione(1991)}]{carcione:1991b}
\bibinfo{author}{Carcione, J.M.}, \bibinfo{year}{1991}.
\newblock \bibinfo{title}{Domain decomposition for wave propagation problems}.
\newblock \bibinfo{journal}{Journal of Scientific Computing}
  \bibinfo{volume}{6}, \bibinfo{pages}{453--472}.
\bibitem[{Chiaia et~al.(2008)Chiaia, Cornetti, Frigo, Cardu, Chiaravalloti
  et~al.}]{cardu:2008}
\bibinfo{author}{Chiaia, B.}, \bibinfo{author}{Cornetti, P.},
  \bibinfo{author}{Frigo, B.}, \bibinfo{author}{Cardu, M.},
  \bibinfo{author}{Chiaravalloti, L.}, et~al., \bibinfo{year}{2008}.
\newblock \bibinfo{title}{A coupled stress and energy criterion for natural and
  artificial triggering of dry snow slab avalanches}, in:
  \bibinfo{booktitle}{The 42nd US Rock Mechanics Symposium (USRMS)},
  \bibinfo{organization}{American Rock Mechanics Association}. pp.
  \bibinfo{pages}{ARMA--08--203}.
\bibitem[{Cooper(1996)}]{cooper:1996}
\bibinfo{author}{Cooper, P.W.}, \bibinfo{year}{1996}.
\newblock \bibinfo{title}{Explosives engineering}.
\newblock \bibinfo{publisher}{VCH New York}.
\bibitem[{Denoth(1989)}]{denoth:1989}
\bibinfo{author}{Denoth, A.}, \bibinfo{year}{1989}.
\newblock \bibinfo{title}{Snow dielectric measurements}.
\newblock \bibinfo{journal}{Advances in Space Research} \bibinfo{volume}{9},
  \bibinfo{pages}{233--243}.
\bibitem[{Deresiewicz and Skalak(1963)}]{deresiewicz:1963}
\bibinfo{author}{Deresiewicz, H.}, \bibinfo{author}{Skalak, R.},
  \bibinfo{year}{1963}.
\newblock \bibinfo{title}{On uniqueness in dynamic poroelasticity}.
\newblock \bibinfo{journal}{Bulletin of the Seismological Society of America}
  \bibinfo{volume}{53}, \bibinfo{pages}{783--788}.
\bibitem[{van~der Eerden and V{\'e}dy(2005)}]{vandererden:2005}
\bibinfo{author}{van~der Eerden, F.}, \bibinfo{author}{V{\'e}dy, E.},
  \bibinfo{year}{2005}.
\newblock \bibinfo{title}{Propagation of shock waves from source to receiver}.
\newblock \bibinfo{journal}{Noise Control Engineering Journal}
  \bibinfo{volume}{53}, \bibinfo{pages}{87--93}.
\bibitem[{Eller and Denoth(1996)}]{eller:1996}
\bibinfo{author}{Eller, H.}, \bibinfo{author}{Denoth, A.},
  \bibinfo{year}{1996}.
\newblock \bibinfo{title}{A capacitive soil moisture sensor}.
\newblock \bibinfo{journal}{Journal of Hydrology} \bibinfo{volume}{185},
  \bibinfo{pages}{137--146}.
\bibitem[{Friedlander(1946)}]{friedlander:1946}
\bibinfo{author}{Friedlander, F.G.}, \bibinfo{year}{1946}.
\newblock \bibinfo{title}{The diffraction of sound pulses. i. diffraction by a
  semi-infinite plane}.
\newblock \bibinfo{journal}{Proceedings of the Royal Society of London. Series
  A. Mathematical and Physical Sciences} \bibinfo{volume}{186},
  \bibinfo{pages}{322--344}.
\bibitem[{Gottlieb et~al.(1982)Gottlieb, Gunzburger and Turkel}]{gottlieb:1982}
\bibinfo{author}{Gottlieb, D.}, \bibinfo{author}{Gunzburger, M.},
  \bibinfo{author}{Turkel, E.}, \bibinfo{year}{1982}.
\newblock \bibinfo{title}{On numerical boundary treatment of hyperbolic systems
  for finite difference and finite element methods}.
\newblock \bibinfo{journal}{SIAM Journal on Numerical Analysis}
  \bibinfo{volume}{19}, \bibinfo{pages}{671--682}.
\bibitem[{Gubler(1976)}]{gubler:1976}
\bibinfo{author}{Gubler, H.}, \bibinfo{year}{1976}.
\newblock \bibinfo{title}{{K{\"u}nstliche Ausl{\"o}sung von Lawinen durch
  Sprengungen}}.
\newblock \bibinfo{type}{Technical Report} \bibinfo{number}{32}. Swiss Federal
  Institute for Snow and Avalanche Research.
\bibitem[{Gubler(1977)}]{gubler:1977}
\bibinfo{author}{Gubler, H.}, \bibinfo{year}{1977}.
\newblock \bibinfo{title}{Artificial release of avalanches by explosives}.
\newblock \bibinfo{journal}{Journal of Glaciology} \bibinfo{volume}{19},
  \bibinfo{pages}{419--429}.
\bibitem[{Ishida(1965)}]{ishida:1965}
\bibinfo{author}{Ishida, T.}, \bibinfo{year}{1965}.
\newblock \bibinfo{title}{Acoustic properties of snow}.
\newblock \bibinfo{journal}{Contributions from the Institute of Low Temperature
  Science} \bibinfo{volume}{20}, \bibinfo{pages}{23--63}.
\bibitem[{Jaeger et~al.(2007)Jaeger, Cook and Zimmerman}]{jaeger:2007}
\bibinfo{author}{Jaeger, J.C.}, \bibinfo{author}{Cook, N.G.},
  \bibinfo{author}{Zimmerman, R.}, \bibinfo{year}{2007}.
\newblock \bibinfo{title}{Fundamentals of rock mechanics}.
\newblock \bibinfo{publisher}{Blackwell Publishing}. \bibinfo{edition}{fourth
  edition} edition.
\bibitem[{Johnson(1982)}]{johnson:1982}
\bibinfo{author}{Johnson, J.B.}, \bibinfo{year}{1982}.
\newblock \bibinfo{title}{On the application of {Biot}'s theory to acoustic
  wave propagation in snow}.
\newblock \bibinfo{journal}{Cold Regions Science and Technology}
  \bibinfo{volume}{6}, \bibinfo{pages}{49--60}.
\bibitem[{Johnson et~al.(1993)Johnson, Brown, Gaffney, Solie, Sturm and
  L.Blaisdell}]{johnson:1993}
\bibinfo{author}{Johnson, J.B.}, \bibinfo{author}{Brown, J.A.},
  \bibinfo{author}{Gaffney, E.S.}, \bibinfo{author}{Solie, D.J.},
  \bibinfo{author}{Sturm, M.A.}, \bibinfo{author}{L.Blaisdell, G.},
  \bibinfo{year}{1993}.
\newblock \bibinfo{title}{Gas gun experiments to determine shock wave behavior
  in snow}.
\newblock \bibinfo{type}{Technical Report} \bibinfo{number}{93-11}. U.S. Army
  Cold Regions Research and Engineering Laboratory. \bibinfo{address}{Hanover
  N.H., U.S.A}.
\bibitem[{Mach(1878)}]{mach:1878}
\bibinfo{author}{Mach, E.}, \bibinfo{year}{1878}.
\newblock \bibinfo{title}{{\"U}ber den {Verlauf} von {Funkenwellen} in der
  {Ebene} und im {Raume}}.
\newblock \bibinfo{journal}{Sitzungsbr. Akad. Wiss. Wien} \bibinfo{volume}{78},
  \bibinfo{pages}{819--838}.
\bibitem[{McClung and Schaerer(2006)}]{mcclung:2006}
\bibinfo{author}{McClung, D.}, \bibinfo{author}{Schaerer, P.},
  \bibinfo{year}{2006}.
\newblock \bibinfo{title}{The Avalanche Handbook}.
\newblock \bibinfo{publisher}{The Mountaineers Books, Seattle WA, U.S.A.}.
  \bibinfo{edition}{3rd} edition.
\bibitem[{Mellor(1973)}]{mellor:1973}
\bibinfo{author}{Mellor, M.}, \bibinfo{year}{1973}.
\newblock \bibinfo{title}{Controlled release of avalanches by explosives}, in:
  \bibinfo{editor}{Perla, R.} (Ed.), \bibinfo{booktitle}{Advances in North
  American avalanche technology: 1972 Symposium}, \bibinfo{publisher}{USDA
  Forest Service, General Technical Report RM-3}. pp. \bibinfo{pages}{37--49}.
\bibitem[{Mellor(1975)}]{mellor:1975}
\bibinfo{author}{Mellor, M.}, \bibinfo{year}{1975}.
\newblock \bibinfo{title}{A review of basic snow mechanics}, in:
  \bibinfo{booktitle}{The International Symposium on Snow Mechanics,
  Grindelwald, Switzerland}, \bibinfo{organization}{IAHS-AISH}. pp.
  \bibinfo{pages}{251--291}.
\bibitem[{Miller et~al.(2011)Miller, Tichota and Adams}]{miller:2011}
\bibinfo{author}{Miller, D.}, \bibinfo{author}{Tichota, R.},
  \bibinfo{author}{Adams, E.}, \bibinfo{year}{2011}.
\newblock \bibinfo{title}{An explicit numerical model for the study of snow's
  response to explosive air blast}.
\newblock \bibinfo{journal}{Cold Regions Science and Technology}
  \bibinfo{volume}{69}, \bibinfo{pages}{156--164}.
\bibitem[{Oura(1952)}]{oura:1952}
\bibinfo{author}{Oura, H.}, \bibinfo{year}{1952}.
\newblock \bibinfo{title}{Reflection of sound at snow surface and mechanism of
  sound propagation in snow}.
\newblock \bibinfo{journal}{Low Temperature Science} \bibinfo{volume}{9},
  \bibinfo{pages}{179--186}.
\bibitem[{Peyret(2013)}]{peyret:2002}
\bibinfo{author}{Peyret, R.}, \bibinfo{year}{2013}.
\newblock \bibinfo{title}{Spectral methods for incompressible viscous flow}.
\newblock \bibinfo{publisher}{Springer Science \& Business Media}.
\bibitem[{Schweizer and Wiesinger(2001)}]{schweizer:2001}
\bibinfo{author}{Schweizer, J.}, \bibinfo{author}{Wiesinger, T.},
  \bibinfo{year}{2001}.
\newblock \bibinfo{title}{Snow profile interpretation for stability
  evaluation}.
\newblock \bibinfo{journal}{Cold Regions Science and Technology}
  \bibinfo{volume}{33}, \bibinfo{pages}{179--188}.
\bibitem[{Shapiro et~al.(1997)Shapiro, Johnson, Sturm and
  Blaisdell}]{shapiro:1997}
\bibinfo{author}{Shapiro, L.H.}, \bibinfo{author}{Johnson, J.B.},
  \bibinfo{author}{Sturm, M.}, \bibinfo{author}{Blaisdell, G.L.},
  \bibinfo{year}{1997}.
\newblock \bibinfo{title}{Snow mechanics - Review of the state of knowledge and
  applications}.
\newblock \bibinfo{type}{Technical Report} \bibinfo{number}{97-3}. US Army Cold
  Regions Research and Engineering Laboratory. \bibinfo{address}{Hanover, N.H.,
  U.S.A.}
\bibitem[{Sidler(2015)}]{sidler:2015}
\bibinfo{author}{Sidler, R.}, \bibinfo{year}{2015}.
\newblock \bibinfo{title}{A porosity-based {Biot} model for acoustic waves in
  snow}.
\newblock \bibinfo{journal}{Journal of Glaciology} \bibinfo{volume}{61},
  \bibinfo{pages}{789--798}.
\bibitem[{Sidler et~al.(2010)Sidler, Carcione and Holliger}]{sidler:2010a}
\bibinfo{author}{Sidler, R.}, \bibinfo{author}{Carcione, J.M.},
  \bibinfo{author}{Holliger, K.}, \bibinfo{year}{2010}.
\newblock \bibinfo{title}{Simulation of surface waves in porous media}.
\newblock \bibinfo{journal}{Geophysical Journal International}
  \bibinfo{volume}{183}, \bibinfo{pages}{820--832}.
\bibitem[{Simenhois and Birkeland(2009)}]{Simenhois:2009}
\bibinfo{author}{Simenhois, R.}, \bibinfo{author}{Birkeland, K.},
  \bibinfo{year}{2009}.
\newblock \bibinfo{title}{The extended column test: Test effectiveness, spatial
  variability, and comparison with the propagation saw test.}
\newblock \bibinfo{journal}{Cold Regions Science and Technology}
  \bibinfo{volume}{59}, \bibinfo{pages}{210--2167}.
\bibitem[{Simioni et~al.(2015)Simioni, Sidler, Dual and
  Schweizer}]{simioni:2015}
\bibinfo{author}{Simioni, S.}, \bibinfo{author}{Sidler, R.},
  \bibinfo{author}{Dual, J.}, \bibinfo{author}{Schweizer, J.},
  \bibinfo{year}{2015}.
\newblock \bibinfo{title}{Field measurements of snowpack response to explosive
  loading}.
\newblock \bibinfo{journal}{Cold Regions Science and Technology}
  \bibinfo{volume}{120}, \bibinfo{pages}{179--190}.
\bibitem[{Sutton et~al.(1981)Sutton, Duennebier and Iwatake}]{sutton:1981}
\bibinfo{author}{Sutton, G.}, \bibinfo{author}{Duennebier, F.},
  \bibinfo{author}{Iwatake, B.}, \bibinfo{year}{1981}.
\newblock \bibinfo{title}{Coupling of ocean bottom seismometers to soft
  bottom}.
\newblock \bibinfo{journal}{Marine geophysical researches} \bibinfo{volume}{5},
  \bibinfo{pages}{35--51}.
\bibitem[{Tessmer et~al.(1992)Tessmer, Kessler, Kosloff and
  Behle}]{tessmer:1992}
\bibinfo{author}{Tessmer, E.}, \bibinfo{author}{Kessler, D.},
  \bibinfo{author}{Kosloff, D.}, \bibinfo{author}{Behle, A.},
  \bibinfo{year}{1992}.
\newblock \bibinfo{title}{Multi-domain chebyshev-fourier method for the
  solution of the equations of motion of dynamic elasticity}.
\newblock \bibinfo{journal}{Journal of Compuational Physics}
  \bibinfo{volume}{100}, \bibinfo{pages}{355--363}.
\bibitem[{Tessmer and Kosloff(1994)}]{tessmer:1994}
\bibinfo{author}{Tessmer, E.}, \bibinfo{author}{Kosloff, D.},
  \bibinfo{year}{1994}.
\newblock \bibinfo{title}{3-d elastic modeling with surface topography by a
  {Chebychev} spectral method}.
\newblock \bibinfo{journal}{Geophysics} \bibinfo{volume}{59},
  \bibinfo{pages}{464--473}.
\bibitem[{Ueland(1993)}]{ueland:1993}
\bibinfo{author}{Ueland, J.}, \bibinfo{year}{1993}.
\newblock \bibinfo{title}{Effects of explosives on the moutain snowpack}, in:
  \bibinfo{booktitle}{Proceedings International Snow Science Workshop,
  Breckenridge, Colorado, U.S.A., 4-8 October 1992},
  \bibinfo{publisher}{Colorado Avalanche Information Center, Denver CO, USA}.
  pp. \bibinfo{pages}{205--213}.
\bibitem[{Wooldridge et~al.(2012)Wooldridge, Hendrikx, Miller and
  Birkeland}]{wooldridge:2012}
\bibinfo{author}{Wooldridge, R.}, \bibinfo{author}{Hendrikx, J.},
  \bibinfo{author}{Miller, D.}, \bibinfo{author}{Birkeland, K.},
  \bibinfo{year}{2012}.
\newblock \bibinfo{title}{The effect of explosives on the physical properties
  of snow}, in: \bibinfo{booktitle}{International Snow Science Workshop ISSW
  2012, Anchorage AK, U.S.A., 16-21 September 2012}, pp.
  \bibinfo{pages}{1033--1039}.

\end{thebibliography}

\end{document}